\documentclass[11pt]{article}
\usepackage[english]{babel}
\usepackage[utf8]{inputenc}
\usepackage[T1]{fontenc}
\usepackage{indentfirst}
\usepackage{amsmath}
\usepackage{amssymb}
\usepackage{eufrak}
\usepackage{graphicx}
\usepackage{psfrag}
\usepackage[font=small]{caption}
\usepackage{subcaption}

\allowhyphens

\newcommand{\valos}{\mathbb{R}}
\newcommand{\complex}{\mathbb{C}}
\newcommand{\eps}{\varepsilon}

\newcommand{\ordo}{\mathcal{O}}

\newtheorem{thm}{Theorem}

\newcommand{\vev}[1]{\left\langle #1 \right\rangle}
\newcommand{\ket}[1]{{\left|#1\right\rangle}}
\newcommand{\bra}[1]{{\left\langle #1\right|}}

\newcommand{\skalarszorzat}[2]{{\langle #1 | #2 \rangle}}

\newtheorem{conj}{Proposition}


\usepackage[sort&compress,numbers]{natbib}

\usepackage{ifpdf}

\ifpdf
\usepackage{epstopdf}
\usepackage[pdftex,colorlinks,urlcolor=blue,citecolor=blue,linkcolor=blue]{hyperref}
\else
\usepackage[hypertex,colorlinks,urlcolor=blue,citecolor=blue,linkcolor=blue]{hyperref}
\fi
\begin{document}
\numberwithin{equation}{section}

\title{On Generalized Gibbs Ensembles with an infinite set
  of conserved charges}
  
\author{B. Pozsgay$^{1,2}$, E. Vernier$^{3}$, M. A. Werner$^{4}$
  \\
  ~\\
  $^1$ Department of Theoretical Physics, Budapest University
of Technology and Sciences,\\
1111 Budapest, Budafoki \'{u}t 8, Hungary\\
$^2$ MTA-BME ``Momentum'' Statistical Field Theory Research
Group, \\
1111 Budapest, Budafoki \'{u}t 8, Hungary\\
$^3$ SISSA and INFN, via Bonomea 265, 34136 Trieste, Italy\\
$^4$ MTA-BME ``Momentum'' Exotic Quantum Phases Research
Group, \\
1111 Budapest, Budafoki \'{u}t 8, Hungary}

\maketitle

\abstract{

We revisit the question of whether and how the steady states arising
after non-equilibrium time evolution in integrable models (and in
particular in the XXZ spin chain)
can be described by the so-called Generalized Gibbs Ensemble (GGE). It
is known that the micro-canonical ensemble built on a complete 
set of charges correctly
describes the long-time limit of local observables,
and recently a canonical ensemble was built by Ilievski et. al. using
particle occupation number operators. Here we provide an alternative
construction 
by considering truncated GGE's (tGGE's) that only include a finite number of
well localized conserved operators. It is shown 
that the tGGE's can approximate the steady states with
arbitrary precision, i.e. all physical observables are exactly reproduced in
the infinite truncation limit. 
In addition, we show that a complete canonical ensemble can in fact be
built in terms of a new (discrete) set of charges built as linear combinations of
the standard ones.  

Our general arguments are applied to
concrete quench situations in the XXZ chain, where the initial states are simple
two-site or four-site product states. Depending on the quench we find
that numerical results for the local
correlators can be obtained with remarkable precision using truncated
GGE's with only 10-100 charges.
}

\section{Introduction}

The equilibration of closed quantum systems has received a lot of
interest recently. One of the main questions is whether the principles
of statistical physics can be derived from the unitary time evolution
of quantum mechanics. Can a big quantum system act as its own thermal
bath, and can the emerging steady states be described by statistical
physical ensembles?
These problems go back to the work of von
Neumann, however, our understanding of thermalization has grown
considerably over the last two decades \cite{Silva-quench-colloquium,Eisert-NatPhys-quench-review}.

An area which has received special interest is the field of integrable
models. These systems possess additional conserved charges on top of
the common conserved quantities. As a result the real-time dynamics of
these models is more restricted and the systems can not equilibrate to
standard statistical physical ensembles. In this respect they are important
exceptions that deserve study on their own. However, their study is
also motivated by their experimental relevance, because they can be
tailored in modern experiments \cite{ultracold-experimental-review,boson-experimental-review,Silva-quench-colloquium,fermi-gases-experiments-review} and they also describe real
world materials \cite{KiserletiOsszefogl-batchelor-foerster1}.

It has been proposed in \cite{rigol-gge} that integrable models
equilibrate to Generalized Gibbs Ensembles (GGE's) which involve all
higher conserved charges of the systems
\cite{rigol-2,rigol-3,rigol-GETH}. The investigation of the GGE has been a
central topic in this field in the last 10 years. Regarding
certain non-interacting 
theories it was established rigorously that they indeed describe the
post-quench steady states (see
\cite{essler-fagotti-quench-review,rigol-quench-review} and references
therein), and signs of the GGE in an 
almost-free model have been
also observed experimentally \cite{gge-experiment1}.

On the other hand, less is known about genuinely interacting theories.
The reason for this is that it is notoriously difficult to compute exact results for the
real time evolution in these models, and typical approximation methods
fail to describe the long time dynamics needed to study
equilibration. Whereas it is generally believed that the GGE gives a
correct description of post-quench steady states, there are a number
of open questions regarding interacting theories. What is the full set
of conserved charges that is required for the GGE? In practice how can one
calculate predictions of the GGE? These questions have been considered
on a case by case basis, and at present no general treatment is available. 

An interacting model that has been investigated thoroughly is the
Heisenberg XXZ model. Its importance lies in the fact that it is
interacting with a tunable coupling, yet it has a relatively simple
solution, and a large literature is already available to the
computation of its correlation functions. Regarding the GGE the first
exact results appeared in 
\cite{sajat-xxz-gge,essler-xxz-gge} which gave predictions for
post-quench correlators using a GGE built on the so-called ultra-local
charges; a closely related work appeared later
\cite{fagotti-collura-essler-calabrese}. Despite the general
expectation at that time
that this ``ultra-local GGE'' should be valid, it was shown in the
parallel works  \cite{JS-oTBA,sajat-oTBA} that this is not true.
These
works used the so-called Quench Action method \cite{quench-action}
which is built on the knowledge of the exact overlaps with the initial
states
\cite{sajat-neel,sajat-Karol,Caux-Neel-overlap1,Caux-Neel-overlap2}.
It was demonstrated in   \cite{JS-oTBA,sajat-oTBA} that the steady
states formed in the quenches of the XXZ chain can be markedly
different from the prediction of the ultra-local GGE.
A complete understanding of this mismatch
was finally given in \cite{JS-CGGE}. It was shown there that an ensemble which
also includes the recently discovered quasi-local charges \cite{enej-review} 
does in fact correctly describe the asymptotic states.

Thus, the general statements of \cite{JS-CGGE} seemed to settle the status
of the GGE in the XXZ chain. However, a surprising and important
additional information was pointed out recently in  \cite{enej-gge}:
it was shown that the canonical ensemble built out of the
aforementioned local and quasi-local charges is generally ill-defined,
because the Lagrange-multipliers associated 
to the higher charges necessarily miss certain information about the
asymptotic states. This is not in contradiction with the results of
\cite{JS-CGGE}, because the construction of \cite{JS-CGGE}, which
correctly describes the steady states, can still
be understood as a micro-canonical ensemble. 
As a solution to this problem, a canonical ensemble
  was constructed in \cite{enej-gge} in terms of the mode occupation
  operators for the physical particles, whose eigenvalues are the
  Bethe root densities.
  These operators are closely related to the
  traditional set of conserved charges, in particular they are defined
  through the same set of commuting transfer matrices.
While such a formulation indeed contains all necessary
information about the steady-states, the role of locality in the GGE
is less transparent. 
The mode occupation number operators
have to be defined using an infinitesimal regulator, which also
guarantees their quasi-locality. The physical predictions of the GGE are
insensitive to this regulator, but the operators themselves that
enter the GGE lose quasi-locality as the regulator is taken to
zero. This is in accordance with free theories, where a GGE can be
constructed using the occupation number operators, which are
given by the necessarily non-local Fourier integrals.

In this work we investigate the GGE from a
different point of view:
we consider truncated GGE's with only a
finite number of charges. This idea goes back to the work
\cite{essler-truncated-gge} where truncated and ``defective'' GGE's were considered
in the Ising chain. Moreover, the paper \cite{sajat-xxz-gge} by one of
the authors also considered truncated GGE's for the XXZ chain
(at that time only the ultra-local charges were
included). The main goal of this work is to show that
a sequence of truncated GGE's (using both the ultra-local and
quasi-local charges) can approximate the post quench steady states with
arbitrary precision.
Going further, we will construct new sets of local and
quasi-local charges from linear combinations of the standard ones,
from which a complete GGE can be defined. It is an
important aspect of this formulation that the Lagrange multipliers
associated to the new charges are well defined physical state functions (``generalized
temperatures'') of the spin chain. 
We will argue that in the infinite truncation limit
  all local observables are reproduced exactly, therefore our
  construction accomplishes the same goal as  \cite{enej-gge},
  although with a different set of charges. 
  In the infinite-truncation limit the range of our operators
  is increased gradually and quasilocality is necessarily lost.  Therefore, this limit
can be understood as the counterpart of the zero-regulator limit in
\cite{enej-gge}.
 
The article is organized as follows. In Section \ref{sec2} we discuss
the general principles of equilibration in quantum systems and the
different statistical ensembles that can describe the long
time behavior in non-integrable and integrable systems.
In Section \ref{micro} we discuss the particular example of the XXZ
spin chain: after a brief introduction to the solution of the model
and the description of the conserved operators we present the
arguments of \cite{JS-CGGE}, which prove that the micro-canonical
ensemble correctly describe the post quench states. In section
\ref{canonical} we present our construction of the truncated GGE's and
prove that they can describe the steady states with arbitrary
precision. Section \ref{examples} deals with concrete 
examples: as initial states we consider the dimer state and two
different four-site product states.
We present numerical evidence for the convergence of the
truncated GGE's to the micro-canonical ensemble. In the case of the four-site states we also present
numerical data for the real-time evolution of local
observables as obtained using the so-called iTEBD method, and a comparison is
made to the predictions of the GGE.

\section{
 Statistical ensembles for post-quench steady states 
 }

\label{sec2}

The general question underlying our work is the possibility to describe the long time behavior of local observables in an isolated quantum system following a quantum quench. 
For a given initial state $\ket{\Psi_0}$ , the time evolution of a local observable $\ordo$ is given by
\begin{equation*}
  \vev{\ordo(t)}=\bra{\Psi_0}e^{iHt} \ordo e^{-iHt} \ket{\Psi_0} \,,
\end{equation*}
and we define the long time limit as
\begin{equation*}
  \bar \ordo=\lim_{T\to\infty} \int_{0}^T dt \vev{\ordo(t)}  \,. 
\end{equation*}
Here the averaging is introduced so that the
limit is well defined even in a finite volume. If one deals with a
global quench in an infinite system, then typically the averaging can be
omitted.

Expanding over a set of eigenstates of the Hamiltonian $H$ and neglecting degeneracies yields the so-called Diagonal Ensemble (DE) description, 
\begin{equation}
  \label{DE}
\bar\ordo=\sum_n |c_n|^2 \bra{n}\ordo\ket{n},\qquad c_n=\skalarszorzat{\Psi_0}{n} \,.
\end{equation}
whose accuracy has been checked for both integrable and non-integrable models \cite{free-gge-2,rigol-eth,eth1,eth2}. 
A natural question to address is therefore that of the relation between this description and the usual ensembles of statistical mechanics.

\subsection{Thermalization, micro-canonical and canonical ensembles}

In equilibrium statistical mechanics, the micro-canonical ensemble average at energy $E_0$ of an observable $\mathcal{O}$ is defined as 
\begin{equation*}
  \vev{\ordo}_{\text{micro}} =\sum'_{n} \bra{n}\ordo\ket{n},
\end{equation*}
where the restricted sum runs over eigenstates whose energies $E_n$ are such that $|E_n-E_0|<\delta$, where the energy window $\delta$ must be small, but still much greater than the mean many-body level spacing. As discussed in \cite{rigol2}, its precise value has no noticeable influence on the resulting value of $\vev{\ordo}_{\text{micro}}$. 
On the other hand, the Gibbs ensemble average at inverse temperature $\beta = 1/T$ is given by
\begin{equation*}
  \vev{\ordo}_{\text{GE}}=\frac{1}{Z} \sum_n  e^{-\beta E_n}  \bra{n}\ordo\ket{n},
\qquad Z=\sum_n e^{-\beta E_n} 
\,. 
\end{equation*}
It is a general prediction of equilibrium statistical mechanics that
in the thermodynamic limit the micro-canonical and canonical ensembles
become equivalent.

In the non-equilibrium context a system is said to
thermalize following a quantum quench if for all local observables  
\begin{equation}
  \label{all-is-equal}
\bar\ordo =   \vev{\ordo}_{\text{micro}}  =  \vev{\ordo}_{\text{GE}} \,,
\end{equation}
where the above relation has to be understood in the thermodynamic limit, and where the micro-canonical average energy $E_0$ and canonical inverse temperature $\beta$ are uniquely fixed by the conservation of energy:
\begin{equation}
  \label{hukkle}
  E_0  = \vev{H}_{\text{GE}}=\bar H=\bra{\Psi_0}H\ket{\Psi_0} \,.
\end{equation}
In establishing \eqref{hukkle} we used the fact that the Hamiltonian is itself a sum of local
observables. Physically, relation \eqref{hukkle} means that the temperature in the 
post-quench steady state is determined by the energy available in the
initial state.

\bigskip

In generic systems, namely in the absence of further conserved
quantities, a widely accepted scenario for explaining thermalization
is that of the Eigenstate Thermalization Hypothesis (ETH)
\cite{eth1,eth2}, according to which the diagonal matrix elements of
local operators in the eigenstates basis vary smoothly with the
energy, and the off-diagonal matrix elements are exponentially small
in the volume. This hypothesis can be used to prove the first equality
in \eqref{all-is-equal} as follows.

If the initial state of the quench problem satisfies the cluster
decomposition principle \cite{spyros-calabrese-gge} (that is,
$\lim_{|x-y|\to \infty} \bra{\Psi_0} \mathcal{O}(x)\mathcal{O}(y)
\ket{\Psi_0} = \bra{\Psi_0} \mathcal{O}(x) \ket{\Psi_0}\bra{\Psi_0}
\mathcal{O}(y) \ket{\Psi_0}$), then the variation of the energy
density becomes 
suppressed as $\ordo(1/\sqrt{L})$:
\begin{equation}
\label{width}
  \Delta \left(\frac{E}{L}\right)
=\frac{1}{L}\sqrt{\bra{\Psi_0}H^2\ket{\Psi_0}-(\bra{\Psi_0}H\ket{\Psi_0})^2}\sim \frac{1}{\sqrt{L}}.
\end{equation}
This means that the eigenstates having a non-negligible overlap with $\ket{\Psi_0}$ will have an energy density very close to the mean value. Together with the ETH this implies that the
Diagonal Ensemble \eqref{DE} and the micro-canonical ensemble yield the
same expectation values.

The second
equality in \eqref{all-is-equal}, namely the equivalence with the Gibbs ensemble, follows from
analogous considerations as those of equilibrium statistical
mechanics, namely the fact that in the thermodynamic limit the energy density
fluctuations of the Gibbs Ensemble become negligible.

\subsection{The case of integrable systems: generalized micro-canonical and generalized Gibbs ensembles}

In integrable models the situation is different due to the existence
of higher conserved charges $\{Q_j\}_{j=1,\dots,N_Q}$. Here $N_Q$ denotes the number of linearly independent local charges, which
typically scales polynomially with the volume of the system and
therefore also goes to infinity in the thermodynamic limit.  

For such models the conservation of the extra charges constrains the post-quench dynamics, therefore equilibration to the standard Gibbs
ensembles cannot be expected. Instead, the so-called Generalized
Gibbs Ensemble (GGE) has been put forward in \cite{rigol-gge}. In the
canonical form the GGE includes all charges with appropriate Lagrange multipliers, and the
ensemble averages are defined as
\begin{equation*}
  \vev{\ordo}_{\text{GGE}}=\frac{1}{Z} \sum_n  e^{-\sum_{j=1,\dots,N_Q}\bra{n}Q_j\ket{n}}  \beta_j\bra{n}\ordo\ket{n},
\qquad Z=\sum_n e^{-\sum_{j=1,\dots,N_Q}\beta_j\bra{n}Q_j\ket{n}},
\end{equation*}
Following \cite{rigol2}, a generalized micro-canonical ensemble is defined as 
\begin{equation*}
  \vev{\ordo}_{\text{Gmicro}} =\sum'_{n} \bra{n}\ordo\ket{n},
\end{equation*}
where the restricted sum runs over all eigenstates whose charges are close to their initial expectation values, 
\begin{equation*}
  |\bra{n}Q_j\ket{n} -  \bra{\Psi_0}Q_j\ket{\Psi_0}|< \delta_{Q_j} \,.
\end{equation*}
These generalized ensembles give a full description of the post-quench steady-state if for all local observables
\begin{equation}
\label{Gtherm}
  \bar \ordo =  \vev{\ordo}_{\text{Gmicro}} = \vev{\ordo}_{\text{GGE}} \,. 
\end{equation}
Regarding the canonical ensemble these relations can be used to fix
the Lagrange multipliers through the relations
\begin{equation}
  \bra{\Psi_0}Q_j\ket{\Psi_0}=\vev{\ordo}_{\text{GGE}}\,.
\end{equation}

As for the case of thermalization discussed in the previous section,
the equality (\ref{Gtherm}) has to be understood in the thermodynamic
limit. The first part of the equality, namely the equivalence between
the long-time limit of local observables and the generalized
micro-canonical average, can be justified similarly as in the thermal
case using the so-called Generalized Eigenstate Hypothesis (GETH)
\cite{rigol-GETH}. The latter, whose validity has been verified in the
case of the XXZ spin chain in \cite{JS-CGGE}, roughly states that the 
diagonal matrix elements of local operators depend only on the
expectation values of the local charges $Q_j$, in other words, that if for two states
all $\bra{n}Q_j\ket{n}$ are close, the mean values $\bra{n}\ordo\ket{n}$ will be
close too. Following the same steps as in the previous paragraph, one
may indeed deduce from there the equivalence  $\bar \ordo =
\vev{\ordo}_{\text{Gmicro}}$, given that the initial state satisfies
the cluster decomposition principle.

Turning to the Generalized Gibbs Ensemble, a number of issues arise. 
First, we stress that in order to define the GGE one has to start
from a finite volume with a finite number of charges, which means in
practice that the equality (\ref{Gtherm}) holds in the limit where
both the volume of the system and the number of charges are taken to
infinity.  We stress that the GGE has strong predictive power even
though it has an infinite set of Lagrange multipliers in the $L\to\infty$
 limit: in finite volume the total number of independent parameters
 scales only polynomially with the volume, whereas the full Hilbert
 space grows exponentially. 

An important concept is the so-called truncated GGE (tGGE)
\cite{essler-truncated-gge,sajat-xxz-gge}. In building the tGGE we
keep a finite number of charges while taking the thermodynamic limit.
The number of charges can be taken to infinity as a second step, and
doing this a number of interesting physical questions can be
addressed. What are the most relevant charges, which need to be added
first the to tGGE? Is it possible to leave out certain charges, can
these ``defective'' tGGE's describe the steady states nevertheless? These
questions have been investigated in the free case
\cite{essler-truncated-gge,sajat-xxz-gge}, and it was found that
the local observables are most influenced by the most local
charges, and that the ``defective'' tGGE's always miss some
information about the steady states.
In the XXZ chain the tGGE was first investigated in
\cite{sajat-xxz-gge}, however, at that time only the ultra-local
charges were added to the ensemble.
To our best knowledge, tGGE's with a complete set of charges
have not yet been investigated in interacting theories.

To conclude this section we put forward that
in the presence of an infinite number of charges 
the equivalence of the micro-canonical
and canonical ensembles is not guaranteed. 
Also, it is not clear whether the infinite number of Lagrange multipliers
are physical state functions in integrable models. In a generic system
with a finite number of charges the $\beta_j$ have fixed physical
values, such as the value of the temperature, or a chemical potential.
However, perhaps surprisingly, this is not necessarily true in an
interacting system with an infinite number of charges. 
In the following sections we will show that the problems can be traced
to inherent differences between finite and infinite dimensional configuration
spaces. The questions will be discussed on the example of the spin chain.

\section{The generalized microcanonical ensemble in the XXZ chain}

\label{micro}

In this section we briefly review the solution of the XXZ spin chain,
the construction of its conserved charges, the GETH and the
generalized microcanonical ensemble in this model. The reader who is 
already familiar with the technical details might skip this section.

The Hamiltonian of the XXZ Heisenberg spin chain is given by
\begin{equation}
  \label{eq:H}
  H_{XXZ}=\sum_{j=1}^L  \left(\sigma_j^x \sigma_{j+1}^x+ \sigma_j^y\sigma_{j+1}^y
+\Delta (\sigma_j^z \sigma_{j+1}^z-1)\right) \,,
\end{equation}
{where the $\sigma_j^{x,y,z}$ are Pauli matrices acting locally on the $j^{\text{th}}$ spin, and where periodic boundary conditions are assumed.} The additive constant is chosen such that the ferromagnetic reference 
state $\ket{0}\equiv \ket{\uparrow\uparrow\dots \uparrow}$ is an eigenstate of $H$ with
zero energy.
The number $\Delta$ is the so-called anisotropy parameter and we will use the
parametrization $\Delta=\cosh(\eta)$. In the present work we will
deal with the so-called massive regime where $\Delta>1$. This
limitation is chosen because of certain mathematical
simplifications that arise in this regime; the details will be given
below. However, we believe that the our main statements are valid
for arbitrary $\Delta$. We note that recently important progress has
been made concerning the GGE in the massless regime with $|\Delta|<1$
\cite{jacopo-massless-1,jacopo-massless-2}. 

The eigenstates of the Heisenberg model can be constructed using the
different forms of the Bethe Ansatz \cite{Korepin-book,Takahashi-book}.
In coordinate Bethe Ansatz
the wave functions can be {parametrized by a set of complex quasi-momenta $\{\lambda_j\}_{j=1,\ldots N}$ (the so-called Bethe roots) as 
\begin{equation}
\label{XXZ-eloallitas}
  \ket{ \{\lambda_j\}_{j=1,\ldots N} }=
\sum_{y_1< y_2<\dots< y_N}
  \phi_N(\{\lambda_j\}_{j=1,\ldots N}|y_1,\dots,y_N)   \sigma^-_{y_1}\dots \sigma^-_{y_N} \ket{0},
\end{equation}
where $\ket{0}$ is the reference state with all spins up introduced above, and  
}
\begin{equation}
\label{xxz-coo-ba}
  \phi_N(\{\lambda_j\}_{j=1,\ldots N}|\{y\})=\sum_{P \in S_N} 
\left[\prod_{1\le m < n \le N}\frac{\sin(\lambda_{P_m}-\lambda_{P_n}+i\eta)}{\sin(\lambda_{P_m}-\lambda_{P_n})}\right]
\left[\prod_{l=1}^N
\left(\frac{\sin(\lambda_{P_l}+i\eta/2)}{\sin(\lambda_{P_l}-i\eta/2)}\right)^{y_l}\right].
\end{equation}
The quasi-momenta $\{\lambda_j\}_{j=1,\ldots N}$ satisfy the so-called Bethe equations, which follow from the periodicity of
the wave function:
\begin{equation}
 \label{XXZBE}
\left(
\frac{\sin(\lambda_j-i\eta/2)}{\sin(\lambda_j+i\eta/2)}
\right)^L \prod_{k\ne j}  
\frac{\sin(\lambda_j-\lambda_k+i\eta)}{\sin(\lambda_j-\lambda_k-i\eta)}=1.
\end{equation}
The associated energy eigenvalues are given by
\begin{equation}
\label{XXZ-energy}
  E=-\sum_j  
\frac{2\sinh^2\eta}{\sin(\lambda_j+i\eta/2)\sin(\lambda_j-i\eta/2)}.
\end{equation}
In the regime $\Delta>1$ the solutions of
the Bethe equations \eqref{XXZBE} are either real or {assemble into regular patterns in the complex plane, the so-called strings \cite{Takahashi-book}}. A $k$-string is a set of rapidities
$\{\lambda_j\}_{j=1\dots k}$ {centered on the real axis} and such that $\lambda_{j+1}-\lambda_j=i\eta+\delta_j$, where the deviations $\delta_j$ are exponentially small in the
volume $L$. It can be seen from the explicit wave function
\eqref{xxz-coo-ba} that the $k$-strings describe bound states of $k$
interacting spin waves. 

In the thermodynamic limit, the centers (real parts) of each type of
  strings become dense on the interval $[-\pi/2, \pi/2]$, and their
  distribution can be described by a set of continuous densities
  $\rho_k(\lambda)$, together with a set of densities for the
  corresponding holes $\rho_{\text{h},k}(\lambda)$. The latter generalize
  to the interacting case the notion of hole excitations in a Fermi
  sea of non-interacting particles. It follows from 
the Bethe equations that {these densities} satisfy \cite{Takahashi-book}
\begin{equation}
\label{rhorh}
\rho_{k}+\rho_{\text{h},k}= \delta_{k,1}s+
d\star \left(
\rho_{\text{h},k-1}+\rho_{\text{h},k+1}
\right),
\end{equation}
where
\begin{equation}
\label{su}
d(\lambda)=1+2\sum_{n=1}^\infty \frac{\cos(2n \lambda)}{\cosh(\eta n)}
\end{equation}
and the convolution of two functions is defined as
\begin{equation}
  (f\star g)(\lambda)=
\int_{-\pi/2}^{\pi/2} \frac{d\omega}{2\pi} f(\lambda-\omega) g(\omega).
\end{equation}

\subsection{{Local and quasi-local conserved charges}}

Conserved charges of the XXZ model can be constructed using the associated family
of commuting transfer matrices. {While the construction of ultra-local charges, that is, which can be written as a sum of densities with support on a finite number of lattice sites, has been known for a long time \cite{Korepin-book}, it has been completed recently by the introduction a set of quasi-local charges satisfying a weaker but mathematically precise form of locality \cite{enej-review}}. Here we just
collect the main results available in the literature and refer the
reader to the earlier works
\cite{prosen-xxx-quasi,JS-CGGE,eric-lorenzo-exact-solutions,lorenzo-eric-rigol}
for more details.

The construction of the transfer matrices starts with the introduction
of the so-called R-matrix, which
acts on $\complex^2\otimes \complex^2$ and is given explicitly as
\begin{equation}
  R(u)= { \frac{1}{\sinh(u + \eta)} }
  \begin{pmatrix}
    \sinh(u+\eta) & & &\\
& \sinh(u)  & \sinh(\eta) & \\
& \sinh(\eta) & \sinh(u) & \\
& & & \sinh(u+\eta)
  \end{pmatrix}.
\label{R}
\end{equation}
Here $u$ is the spectral parameter.

The fundamental monodromy matrix is constructed as
\begin{equation*}
  \tau(u)=L_M(u)\dots L_1(u),
\end{equation*}
where $L_j(u)$ are local Lax-operators given by
\begin{equation*}
  L_j(u)=R_{0j}(u),
\end{equation*}
and the index 0  stands for an auxiliary spin space.
The transfer matrix is given by the trace over the auxiliary space of the fundamental monodromy matrix, namely
\begin{equation}
  \label{fundat}
  T(u)=\text{Tr}_0 \tau(u) \,.
\end{equation}

It is useful to introduce the higher spin monodromy matrices as
\begin{equation*}
  \tau_s(u)=L^s_M(u)\dots L^s_1(u),
\end{equation*}
where $L^s_j(u)$ are the so-called fused Lax-matrices that act on the tensor product of
the physical spin at site $j$ and an $s+1$ dimensional auxiliary
space {(here and in the following we stress that $s$ stands for twice the value of the auxiliary spin)}. 
Explicitly, $L^s_j(u)$ can be written as a $2 \times 2$ matrix acting
  on the $j$-th physical spin, such that its entries are expressed in terms of the spin-$s/2$ generators $S_z^s, S_+^s, S_-^s$ acting on the auxiliary space \cite{JS-CGGE,eric-lorenzo-exact-solutions}, namely 
\begin{equation}
L^s_j(u) =  \frac{1}{\sinh \left(u +  (s+1) \frac{\eta}{2} \right)} \left( 
\begin{array}{cc}
\sinh\left(u + \eta/2 + \eta{S_z^s} \right) & S_-^s \\
S_+^s & \sinh\left(u + \eta/2 - \eta{S_z^s} \right) 
\end{array}
\right) \,.
\end{equation}

Higher spin transfer matrices are then defined as the trace over the
auxiliary space of the monodromy matrices, 
\begin{equation*}
  T_s(u)=\text{Tr}_0 \tau_s(u) \,.
\end{equation*}
In these notations $s=1$ corresponds to the fundamental transfer
matrix \eqref{fundat}.

It can be proven using the famous Yang-Baxter equation
\cite{Yang-nested,Baxter-book}
that the transfer matrices commute for all auxiliary spin and all
rapidity parameters:
\begin{equation*}
  [T_s(u),T_t(v)]=0.
\end{equation*}
This property can be used to define the conserved charges of the
model. However, the discussion of the charges depends on the spin of
the auxiliary transfer matrix.

The fundamental transfer matrix \eqref{fundat} gives rise to the
so-called ultra-local charges of the model, which can be defined as follows.
First of all it is easy to see that
\begin{equation*}
  T(0)=U,
\end{equation*}
with $U$ being the translation operator on the chain  and it can be
considered as the first conserved charge: $U=e^{iQ_1}$, where $Q_1$ is
the momentum operator. The other charges can be defined as logarithmic
derivatives of the transfer matrix at $u=0$:
\begin{equation}
\label{Qjdef}
  Q_j= i \left(\partial_\lambda \right)^{j-1} \left. \log T(- i \lambda) \right|_{\lambda=0}.
\end{equation}
It was shown in \cite{Luscher-conserved} that the $Q_j$ defined this way are local
in the sense that they are given
as sums of products of spin variables such that they only span
a finite segment of the chain of length $j$. 
We note that using the normalizations above the second conserved
charge is 
\begin{equation}
  \label{Q2ident}
 Q_2=\frac{1}{2\sinh\eta} H_{XXZ}.
\end{equation}

The higher spin transfer matrices $T_s(u)$ can be used to define
an additional set of charges through
\begin{equation}
  \label{Xsdef}
  X_{s}(\lambda)= \frac{1}{i} \partial_\lambda \log {T_s(- i \lambda)} \,,
\end{equation}
where
\begin{equation}
  \label{Qsj}
Q_{s,j}= - \left.(\partial_\lambda)^{j-1} X_s(\lambda)\right|_{\lambda=0}\,.
\end{equation}
In order to have unified notation we identify $Q_{1,j}=Q_j$ with $Q_j$
defined in \eqref{Qjdef}.
It follows from the commutativity of the transfer matrices and the
identification \eqref{Q2ident} that all $Q_{s,j}$ are conserved under
the time evolution generated by $H_{XXZ}$.

It is important that neither the rapidity{-}dependent $X_s(\lambda)$, nor the operators in the discrete set
$Q_{s,j}$ are local in the usual sense, namely they involve contributions of arbitrarily large support on the spin chain.  
However, it can be shown that for all $\lambda$  within the physical
strip $|\Im(\lambda)|<\eta/2$, the $X_s(\lambda)$ satisfy a weaker
form of locality referred to as quasi-locality
\cite{prosen-xxx-quasi}, a mathematical definition of which can be
found for instance in \cite{enej-review}.  
It is then straightforward to observe that the charges $Q_{s,j}$ defined in (\ref{Qsj}) are themselves quasi-local. 

Most importantly, they are extensive.

The eigenvalues of the charges on Bethe states can be derived simply
from the transfer matrix eigenvalues. However, strikingly simple
relations have been found in  \cite{JS-CGGE} in the thermodynamic
limit. First, for a given Bethe state let us define the generating functions 
\begin{equation}
\label{Gs}
G_{s}(\lambda)=-\frac{1}{L}\vev{X_s(\lambda)}=
\frac{1}{L}\sum_{j=1}^\infty \frac{\lambda^{j-1}}{(j-1)!}\vev{Q_{s,j}} \,,
\end{equation}
{where $\vev{\cdot}$ denotes eigenvalues on the corresponding Bethe state.}
It has been proven in   \cite{JS-CGGE} that for each spin $s$ the
generating function is related to the hole densities of the $s$-strings
through 
\begin{equation}
\label{stringcharge}
  d\star(-a_s+\rho_{\text{h},s})=G_s,
\end{equation}
Here $a_s$ is a simple function given by
\begin{equation}
  \label{ajdef2}
  a_s(\lambda)=\frac{\sinh(s\eta)}{\sin(\lambda+i s\eta/2)\sin(\lambda-i s\eta/2)}=
\frac{2\sinh(s\eta)}{\cosh(s\eta)-\cos(2\lambda)} \,.
\end{equation}
The relation \eqref{stringcharge} has been termed ''string-charge
duality'' in \cite{jacopo-massless-1}, because it connects the string
index to the spin of the quasi-local charges
\footnote{We note that the idea that
the spin of a fused transfer matrix can correspond to a string index
already appeared earlier in the literature, for example in
\cite{TBA-QTM-Kluemper-Takahashi} where the equivalence of the
so-called Quantum Transfer Matrix and Thermodynamic Bethe Ansatz
methods was studied.}.

\subsection{{The generalized microcanonical ensemble}}

The convolution in \eqref{stringcharge} can be inverted explicitly. If there are no poles of $G_s$
in the physical strip then we have
\begin{equation}
\label{rhohfromG}
  -a_s(\lambda)+\rho_{\text{h},s}(\lambda)=G_s(\lambda+i\eta/2)+G_s(\lambda-i\eta/2)\,.
\end{equation}
Using \eqref{rhorh} we obtain
\begin{equation}
  \label{rhorfromG}
  \rho_{s}(\lambda)=-  G_s(\lambda+i\eta/2)-G_s(\lambda-i\eta/2)
  +G_{s+1}(\lambda)+G_{s-1}(\lambda)\,.
\end{equation}
The relations \eqref{rhohfromG}-\eqref{rhorfromG} imply that the
expectation values of the conserved charges completely determine the
Bethe states. In fact, it was shown in \cite{JS-CGGE} that these
relations can be used to prove that the {generalized microcanonical ensemble} gives a
correct description of the steady states after quantum quenches. For
the sake of completeness we reproduce this argument.

If the initial state $\ket{\Psi_0}$ is such that the variance of the
conserved charges behaves as
\begin{equation}
  \label{Qclust}
  \bra{\Psi_0}Q_{s,j}^2\ket{\Psi_0}- \bra{\Psi_0}Q_{s,j}\ket{\Psi_0}^2\sim\ordo(L)\,,
\end{equation}
then conservation of all $Q_{s,j}$ implies that only those Bethe
states can have a non-negligible overlap with $\ket{\Psi_0}$ whose
hole and root densities are calculated from
\eqref{rhohfromG}-\eqref{rhorfromG} with
\begin{equation*}
G_s(\lambda)= {- \frac{1}{L}} \bra{\Psi_0}X_s(\lambda)\ket{\Psi_0}\,.
\end{equation*}
Relation \eqref{Qclust} is guaranteed for states satisfying the
cluster decomposition principle.
On the other hand, it has been known since the early research on
Algebraic Bethe Ansatz that the
root and hole densities completely determine the expectation values of
local observables \cite{Korepin-book}. Also, an explicit and efficient
way to
compute so-called factorized formulas for the correlators has been
presented in \cite{sajat-corr}.

It follows from the above that the first assumption of the Generalized
Eigenstate Thermalization Hypothesis is valid in the XXZ
spin chain: the expectation values of local observables only depend on
the macroscopic values of the complete set of conserved charges. The
second assumption concerning the sufficiently fast decay of the off-diagonal matrix
elements has to be investigated separately, and there are cases when
it is indeed not satisfied \footnote{Although not treated in this
  paper, the off-diagonal matrix elements are responsible for the lack
of symmetry restoration in certain cases when the initial state does
not share the symmetries of the Hamiltonian. }. However, in those
cases when the second assumption also holds the GETH implies
immediately that the microcanonical GGE using the full set of
conserved charges is indeed correct.

It is interesting to note that the overall magnetization is completely
determined by the set of charges, which follows directly from
\eqref{rhorfromG}. Therefore the conserved operator $S^z$ is not
independent in this approach.

\section{{Building the canonical GGE in the XXZ chain}}

\label{canonical}

In this section we move to the core of the present work, which is to investigate whether the steady-states produced by quantum quenches can be described using a (canonical) Generalized Gibbs Ensemble. 
 In the previous section we have already seen that the
generalized microcanonical formulation is correct, since the expectation values of local and quasi-local conserved charges in
the initial state $\ket{\Psi_0}$ 
completely determine all
root densities $\rho_k(\lambda)$, which are the only physical information
needed to describe the local observables in the thermodynamic limit.
Therefore, the remaining question is 
whether there is a canonical ensemble which produces Bethe states with the
desired root densities $\rho_k(\lambda)$.

\subsection{The canonical GGE}

We start with the most natural construction of a canonical GGE in
terms of the local and quasi-local charges $Q_{s,j}$ introduced in the
previous section. For the purpose of all computations it is in fact necessary to start from a finite number of charges, therefore we define a sequence of truncated GGE (tGGE) density matrices as
\begin{equation}
  \label{seriesdef}
  \varrho_n=\frac{1}{Z}\exp\left(
    -\sum_{s=1}^{N_s(n)}\sum_{j=1}^{N_d(n)} \beta_{s,j}Q_{s,j}
    \right) \,,
  \end{equation}
  where the parameters $N_s(n)$ and $N_d(n)$ are truncation numbers, which depend on the sequence index $n$, and $Z$ always denotes the partition function such
that $\text{Tr} \varrho_n=1$. The parameters  $N_s(n)$ and $N_d(n)$ could in principle depend on the volume, however, in the present work we follow the
approach that the thermodynamic limit is performed first with fixed
truncation numbers.
In order to capture all information of the steady states eventually one has to add an
infinite number of charges, therefore the correct density matrix should be reached in the $n\to\infty$ limit, where both $N_{s,d}(n) \to \infty$, even though the specific way of approaching infinity is not specified at this stage.

\begin{figure}
  \centering
\includegraphics[scale=0.45]{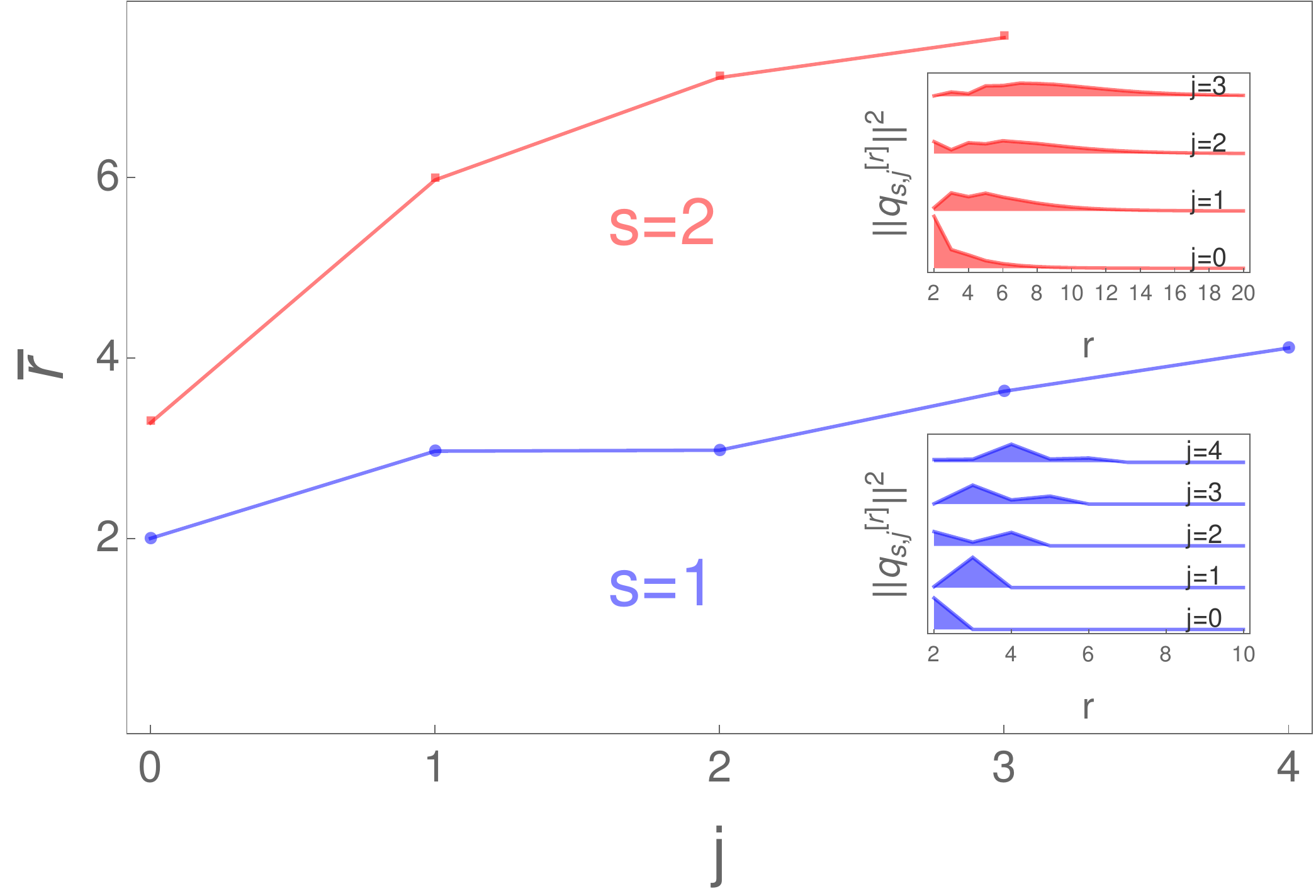}
  \caption{
Average range (\ref{raverage}) of the charges $Q_{s,j}$ for $s=1$ and $s=2$ respectively, plotted as a function of the derivative index $j$ for the first few values of $j$. The anisotropy parameter of the model is taken to be $\eta = 1$. In inset we have represented the Hilbert-Schmidt norms $|| q_{s,j}^{[r]}  ||_{\rm HS}^2$ of the corresponding densities as a function of the range $r$. 
}
  \label{fig:norms}
\end{figure}

The truncated GGE with only the ultra-local
charges ($N_s=1$) was first studied in \cite{sajat-xxz-gge}. In that
case the choice for increasing $N_d$ gradually is physically
motivated: higher charges are less and less local, and it was already
expected in \cite{essler-truncated-gge} that they would have
decreasing influence on
short-range observables. In fact, this was observed in
\cite{sajat-xxz-gge}, even though a general theoretical proof for this
behaviour is still missing.
For the quasi-local charges it is not immediately evident why one
should add the charges with small derivative indices first, since for a given value of the auxiliary spin $s>1$ all these charges are sums of densities with arbitrarily large range. More precisely one can write 
\begin{equation}
Q_{s,j} = \sum_{i}  \mathcal{P}_i \left( \sum_{r \geq 2} q_{s,j}^{[r]} \right)  \,,
\end{equation}
where the first sum runs over the lattice sites and the operator $\mathcal{P}_i\left(\ldots \right)$ denotes a translation by $i$ lattice sites, while the densities $q_{s,j}^{[r]}$ have a non-trivial action on $r$ consecutive lattice sites only and act as identity on the rest of the chain. 
For a given $s > 1$ the norm of these densities is known \cite{enej-review} to decay exponentially with $r$, with a rate independent of the derivative index $j$. Despite this fact, we can argue that, as the ultra-local charges corresponding to $s=1$, the quasi-local charges become ``less and less local'' when $j$ is increased. This can be seen, for instance, by looking at the average ranges $\bar{r}_{s,j}$ of these charges, which we define as (see also \cite{eric-axel})
\begin{equation}
\bar{r}_{s,j} =  \frac{\sum_{r \geq 2}  r || q_{s,j}^{[r]}  ||_{\rm HS}^2}{\sum_{r \geq 2}   || q_{s,j}^{[r]}  ||_{\rm HS}^2}  \,,
\label{raverage}
\end{equation}
where $|| \cdot ||_{\rm HS}$ denotes the Hilbert-Schmidt norm. 
On figure \ref{fig:norms}, we report the values of $\bar{r}_{s,j}$ for $s=1$ and $s=2$, as a function of the derivative index $j$.  Both in the ultra-local ($s=1$) and quasi-local ($s=2$) cases, these are indeed seen to increase with $j$. 
This motivates building a truncated GGE starting from the smallest values of $j$, and the usefulness of this approach will further become evident 
 from our calculations presented below.

\subsection{Review of the results of Ilievski {\it et. al.}}

An alternative way towards a canonical GGE is considered in
\cite{enej-gge}, where the full rapidity
dependent family of quasi-local charges $\{X_s(u)\}$ is added directly to the GGE density matrix. In this case
the Lagrange-multipliers also become functions of the rapidity
variable $u$.
The density matrices can be written as
\begin{equation}
  \label{intGGE}
  \varrho=\frac{1}{Z}
  \exp\left(-\sum_{s=1}^{N_s} \int_{-\pi/2}^{\pi/2} du\ \lambda_s(u) X_s(u)\right)\,.
\end{equation}

These densities matrices were analyzed in \cite{enej-gge} using
the Thermodynamic Bethe Ansatz (TBA) method 
\cite{Takahashi-book}. 
Introducing the standard functions
$Y_s(u)=\rho_{h,s}(u)/\rho_{s}(u)$ which can be interpreted as the
analogue of the (inverse) filling fractions of free theories, the
resulting Bethe states were shown to satisfy
\begin{equation}
  \label{CGGE-enej}
  \begin{split}
  \log(Y_s)&=\lambda_s\star d    +d\star
  \left(\log(1+Y_{s+1})+\log(1+Y_{s-1})
  \right),
        \end{split}
      \end{equation}

It was shown in \cite{enej-gge} that the solution of
\eqref{CGGE-enej} satisfies to so-called modified $Y$-system relations
\begin{equation}
  \label{slh}
  Y_s^+Y_s^-=e^{\lambda_s} (1+Y_{s-1})(1+Y_{s+1}),
\end{equation}
where we introduced the notation $f^{\pm}(u)=\lim_{\eps\to 0}f(u\pm i(\eta/2-\eps))$. This
relation is easily obtained using the pseudo-inverse of the convolution
with $d$, which can be defined as \cite{enej-gge}
\begin{equation}
  (d^{-1} f)(u)=f^+(u)+f^-(u)\,.
\end{equation}
It follows from the Fourier decomposition of $u$ that $d^{-1} d\star
f=f$ for all functions and indeed this leads to \eqref{slh}
immediately. However, $d\star d^{-1}f$ is only equal to $f$ if it does
not have any poles in the physical strip.

As a consequence of \eqref{slh}, it was argued in \cite{enej-gge} that the canonical GGE \eqref{intGGE}
can not describe the post-quench steady states. Probably the simplest
way to see this is to consider 
initial states that are
products of local two-site states, in which cases
exact
solutions for the $Y$-functions are known 
\cite{js-hosszu-kvencs,sajat-QA-GGE-hosszu-cikk,eric-lorenzo-exact-solutions}.
It has been demonstrated earlier and finally proven in
\cite{sajat-Loschm} that they satisfy the original $Y$-system  
\begin{equation}
  \label{Ysys}
  Y_s^+Y_s^-= (1+Y_{s-1})(1+Y_{s+1})\,.
\end{equation}
It follows that if a GGE of the form \eqref{intGGE} would require 
$\lambda_s\equiv 0$ in order to describe these states, which is an apparent contradiction.

It has been pointed out in \cite{enej-gge} that the reason for this
mismatch is
that the Lagrange-multipliers $\lambda_s(u)$ necessarily miss certain
information which can be encoded in the ensembles that use the root
density operators directly:
\begin{equation}
  \label{rhorho}
\varrho=  \frac{1}{Z}
  \exp\left(-\sum_{s=1}^{\infty} \int_{-\pi/2}^{\pi/2} du\ \mu_s(u)
    \hat \rho_s(u)\right)\,, 
\end{equation}
where
\begin{equation}
  \label{rhofromX}
  \hat \rho_{s}(u)= \lim_{\eps\to 0^+}
\left[
  X_s(u+i\eta/2- i \eps)+X_s(u-i\eta/2 + i \eps) - X_{s+1}(u) - X_{s-1}(u)\right]\,.
\end{equation}
It follows from \eqref{stringcharge} that the eigenvalues of the
operators $\hat \rho_s(u)$ are the Bethe root densities $\rho_s(u)$,
given that the functions $X_s(u)$ don't have poles in the physical strip.
Cosidering the density matrices \eqref{rhorho} in the Bethe Ansatz
basis it is evident that they can produce arbitrary post-quench
states, because the Lagrange-multipliers are coupled to the root
densities directly. 

The infinitesimal regulator in \eqref{rhofromX} plays an important
role: the operators $\rho_s(u)$ are well-defined and quasi-local only for
 $\eps>0$. On the other hand, quasi-locality is lost in the $\eps\to 
0$ limit.
This is in accordance with free theories, where the particle
number operators are Fourier modes of the free fields, which are
inherently non-local quantities.
The operators $\hat \rho_s(u)$  provide an exact resolution of the particle
densities in rapidity space, which
requires infinite integration in real space. On the other hand,
restricting the range of the operators to a finite length $L$ would result
in smeared densities with precision $\eps=1/L$.

\bigskip

Coming back to the truncated GGE's introduced in the previous
paragraph, it is tempting to identify the integral form $\int
\lambda_s(u) X_s(u)$ with the $N_d\to\infty$ limit of the sequence
\eqref{seriesdef}.
However, there are a number of subtle issues regarding this identification. 
Assuming that the identification can be established, it
follows from the results of \cite{enej-gge} that the tGGE of the form \eqref{seriesdef}
can not describe the post-quench
steady-states. This motivate us to relax the requirement that the 
Lagrange multipliers $\beta_{s,j}$ should have
fixed values independent of the truncation index.
 In the next section we will show that if we allow
for {\it varying}
$\beta_{s,j}(n)$ as a function of the truncation index $n$, then this
way we can describe all Bethe root distributions. On the other
hand, if the $\beta_{s,j}$ are allowed to vary with $n$, the resulting 
 tGGE can not be identified with \eqref{intGGE} anymore.

The original construction \eqref{seriesdef} reflects physical intuition from standard non-integrable
systems, where the finite number of Lagrange multipliers are actual state functions
and have well defined physical
values. In integrable models  the situation is different due to the
infinite number of charges.
We will show 
that the unphysical nature of the varying $\beta_{s,j}$ can be
 be remedied by constructing a new set of charges, so that the associated
Lagrange multipliers become well defined state
functions. Moreover, we expect that our construction
  is an alternative way towards the $\eps\to 0$ limit of the GGE
  density matrix \eqref{rhorho}.

\subsection{Construction of truncated GGEs}

According to the discussion above we  consider sequences of
tGGE's where each $\beta_{s,j}$ is allowed to depend on the 
truncation index $n$, namely
\begin{equation}
  \label{seriesdef2}
  \varrho_n=\frac{1}{Z}\exp\left(
    -\sum_{s=1}^{N_s(n)}\sum_{j=1}^{N_d(n)} \beta_{s,j}(n)Q_{s,j}
    \right) \,.
\end{equation}

There are two approaches to fixing the Lagrange multipliers.
One possibility is to require that at each truncation step the
operators that are already added will acquire the correct mean values dictated by
the initial state:
\begin{equation}
  \label{req123}
  \text{Tr}\left(\varrho_n Q_{s,j} \right)=\bra{\Psi_0}Q_{s,j}\ket{\Psi_0},\qquad
  s=1,\dots,N_s(n), \quad j=1,\dots,N_d(n)\,.
\end{equation}
Regarding the ultra-local charges this approach was considered in
\cite{sajat-xxz-gge}.
The alternative approach is to relax \eqref{req123} and only require that the
charges obtain the correct values in the infinite truncation limit:
\begin{equation}
  \label{req1234}
\lim_{n\to\infty}  \text{Tr}\left(\varrho_n Q_{s,j} \right)=\bra{\Psi_0}Q_{s,j}\ket{\Psi_0},\qquad
  s=1,\dots,\infty \quad j=1,\dots,\infty\,.
\end{equation}
It follows from the string-charge duality \eqref{stringcharge} that
this merely means that in the $n\to\infty$ limit the correct $\rho_k(u)$ Bethe
root distributions are reproduced.

In the present work we will follow the second approach, because it
enables us to perform certain mathematical steps more easily. However,
we stress that from a physical point of view the two approaches can be
considered equivalent, because eventually we will be interested in the
physical quantities in the $n\to\infty$ limit, when both the local and
quasi-local quantities will take their correct physical values.

Using once again a TBA strategy, the density matrices of the type \eqref{seriesdef2} produce
states satisfying
\begin{equation}
  \label{CGGE-TBA3}
  \begin{split}
  \log(Y_s)&=    -\delta_{s\le N_s}\sum_{k=1}^{N_d}\beta_{s,k} d^{(k)}+d\star
  \left(\log(1+Y_{s+1})+\log(1+Y_{s-1})
  \right),
        \end{split}
      \end{equation}
      where
      \begin{equation*}
        d^{(k)}(u)=\left(\frac{d}{du}\right)^{k-1} d(u)\,.
      \end{equation*}

Note that in both TBA equations \eqref{CGGE-enej} and \eqref{CGGE-TBA3}, the spin index of
the Lagrange multipliers is identified with the string index of the
$Y$-function, and this is analogous to the relation \eqref{stringcharge}.
Moreover, the structure of the individual source terms reflects the
definition of the charges entering the GGE's.

Our claim now is that any post-quench steady-state can be described
this way. More precisely, we will argue that the following proposition is
true: 
\begin{conj}
  \label{conj:corr}
  For any initial state $\ket{\Psi_0}$ with well defined charges
 there exists a sequence
  of tGGE's
\begin{equation}
  \label{eq:CGGE2b}
  \varrho_n=\frac{1}{Z}\exp\left(
    -\sum_{s=1}^{N_s(n)}\sum_{j=1}^{N_d(n)} \beta_{s,j}(n)Q_{s,j}
    \right)\,,
  \end{equation}
  such that the local observables in the post quench steady state will be given
  by the infinite truncation limit
\begin{equation*}
\lim_{T\to\infty} \int_{0}^T dt \vev{\ordo(t)}
  =\lim_{n\to\infty} \vev{\varrho_n \ordo}\,.
\end{equation*}
\end{conj}
The rest of this section will be devoted to the proof of this
proposition, and to analyze the corollaries of the proof.

The mean values of local correlators in
the Bethe states only depend on the root densities \cite{Korepin-book,sajat-corr}, therefore it is
enough to prove the following:
\begin{conj}
  For any Bethe root distribution $\{\rho_k(u)\}$ there exists a sequence
  of tGGE's
\begin{equation}
  \label{eq:CGGE2}
  \varrho_n=\frac{1}{Z}\exp\left(
    -\sum_{s=1}^{N_s(n)}\sum_{j=1}^{N_d(n)} \beta_{s,j}(n)Q_{s,j}
    \right)\,,
  \end{equation}
  such that, for real $u$,
  \begin{equation}
    \label{limpupu}
    \lim_{n\to\infty} \rho^n_k(u)=\rho_k(u),
  \end{equation}
where $\rho^n_k(u)$ is the sequence of root densities  obtained
from $\varrho_n$. In \eqref{limpupu} the convergence is understood in
$L_2$ norm, but it is not necessarily uniform in the 
string index $k$.
\end{conj}

The main idea of the proof of this proposition is to write down a
fictitious TBA satisfied by the physical root densities and to
approximate the fictitious source terms using the discrete set of
charges.
To this end we write down a TBA-like equation
and look for the apparent sources:
\begin{equation}
  \label{fictitious}
  \log \frac{\rho_{h,s}(u)}{\rho_s(u)}=\
 \delta_s(u)+d\star
  \left[\log\left(1+ \frac{\rho_{h,s-1}(u)}{\rho_{s-1}(u)}\right)
+\log\left(1+ \frac{\rho_{h,s+1}(u)}{\rho_{s+1}(u)}\right)\right]\,.
\end{equation}
Here the functions $\delta_s(u)$
are merely some remainder functions
determined from the physical root densities, which in turn are
determined from the charges of the initial state through relation \eqref{stringcharge}.

The idea is to construct a series of TBA equations of the form
\eqref{CGGE-TBA3} which will approximate the source terms $\delta_s(u)$ with
arbitrary precision. This can be achieved by noting that the source
terms in \eqref{CGGE-TBA3} involve linear combinations of all the
derivatives of the function $d(u)$. For such situations the following
theorem holds:
\begin{thm}
  \label{remling}
  Let $d(u)\in L_2([-\pi/2,\pi/2])$ such that
  all of its Fourier components are non-zero. Let $D$ be the set of functions
  formed from $d(u)$ and its derivatives:
  \begin{equation*}
    D=\{d(u),d'(u),d''(u),\dots\}\,.
  \end{equation*}
  Then the linear span of $D$ is dense in
  $L_2([-\pi/2,\pi/2])$.
  In other words, for any function
$f\in L_2([-\pi/2,\pi/2])$ there is a sequence
\begin{equation}
  \label{apprseq}
  f_n=\sum_{j=1}^n \alpha_j(n) d^{(j-1)}\,,
\end{equation}
such that $\lim_{n\to\infty} f_n=f$. 
\end{thm}
The proof to this theorem was given by Christian Remling at the
website Math.Stackexchange \cite{876492}. The technical details of the
proof are not relevant to the content of this paper, therefore we do not
reproduce the proof here. However, we stress that it is essential that
all Fourier components of $d(u)$ are non-zero: any missing Fourier
mode would be outside the linear span of the set $D$.

It is important to clarify in what sense $D$ can or can not be considered a basis
in $L_2([-\pi/2,\pi/2])$. One of the common concepts for infinite
dimensional spaces is the Schauder
basis, whose definition is the following. A set $\{b_j\}$ of vectors
of a Banach space is a Schauder basis, if for every vector $v$ there
is a unique set of coefficients $\{\alpha_j\}$ such that 
\begin{equation*}
v= \lim_{n\to\infty}\sum_{j=1}^n \alpha_j b_j\,.
\end{equation*}
In the present case it can be shown that the set $D$ is not a Schauder
basis of $L_2([-\pi/2,\pi/2])$.
In fact, the coefficients of the approximating sequence \eqref{apprseq}
typically vary with $n$. To prove that $D$ can not be a basis it is enough
to show that one of its elements can be approximated with arbitrary
precision using the remaining functions. This is easily seen by
considering the set $D_2=\{d'',d''',d'''',\dots\}$, which is of the
same type as $D$ except that the constant terms of all functions are zero.
It follows that any function whose integral is zero can be
approximated using $D_2$. We can apply this 
argument to $d'$. It follows that the expansion of a function
$f$ using the elements of $D$ can not be unique, therefore $D$ can not
be a Schauder basis. This has interesting physical consequences, which
will be discussed below.

Returning to the construction of the approximating TBA, Theorem
\ref{remling} can be applied directly to the approximation of the 
individual source terms $\delta_s(u)$ in \eqref{CGGE-TBA3}. This means that 
it is possible to obtain a sequence of sets of
Lagrange-multipliers, such that for each $s$
\begin{equation}
  \label{eloall}
  \delta_s(u)=\lim_{n\to\infty}\left(  \sum_{j=1}^n \beta_{s,j}(n) d^{(j-1)}(u)\right)\,.
\end{equation}
In the tGGE's \eqref{seriesdef2} there are always a finite number of
charges included, so it is not possible to approximate all source
terms at the same time. Instead, both $N_s$ and $N_d$ have to be
increased gradually.
This can be performed in many ways, the simplest one is probably to
set $N_s=N_d=n$. Doing this we only approximate the first $n$ nodes of
the TBA equations, and the remaining ones will be left with vanishing
source, as in the case of the thermal TBA. One might ask whether this
approximation scheme is sufficient to obtain correct values for the
physical observables.

It is a general
experience with the TBA that the influence of the sources for the higher strings 
on the root densities of the smaller strings is small. Also, the root
densities of the higher strings typically don't change the values
of the local correlators; in fact all local correlators can be
expressed using the 1-string nodes of certain auxiliary functions only
\cite{sajat-corr,sajat-QA-GGE-hosszu-cikk}. It follows that a gradual
increase of $N_s$ and $N_d$ will result in the correct values of the
local correlators, even if the convergence of the sources and root
densities is not uniform in the string index.
At present we can not make this argument mathematically precise,
therefore we also demonstrate this approximation procedure on concrete
examples in the next section.

An important technical issue is the asymptotic behavior of the
TBA for large string indices. It is known that depending on the
situation the asymptotics can be very different
\cite{Takahashi-book,JS-CGGE,js-hosszu-kvencs,eric-lorenzo-exact-solutions,sajat-Loschm}.
However, it is expected that if all low lying TBA nodes have the
correct non-trivial source terms, then a large class of asymptotic
behaviors will lead to the same root densities in the infinite
truncation limit.

In the remainder of this section we present a few remarks and
questions regarding our construction of the tGGE's.

\begin{itemize}

\item It is interesting to consider the Lagrange multipliers with the lowest
derivative index. To do this we integrate equation \eqref{eloall} and obtain
\begin{equation}
  \lim_{n\to\infty} \beta_{s,1}(n) =
   \frac{1}{\pi}\int_{-\pi/2}^{\pi/2} du\ \delta_s(u)\,.
\end{equation}
This means that all $\beta_{s,1}$ have well defined physical values determined by the
post-quench steady state, and therefore they could be considered state
functions. Quite interestingly this includes the parameter
$\beta_{1,1}$ which is the coefficient of the Hamiltonian in the GGE,
and therefore can still be identified with the inverse temperature.
In our numerical investigations (to be presented in section
\ref{examples})  we found that all other $\beta_{s,j}$ typically 
oscillate and indeed only $\beta_{s,1}$ tend to fixed values.
We will however see in the next subsection that linear combinations of
the  $\beta_{s,j}$'s can be formed, which have well defined values
 and can therefore be interpreted as ``generalized
inverse temperatures''. 

\item It is worthwhile to consider the $Y$-system equations. The TBA
\eqref{CGGE-TBA3} following from the tGGE's necessarily leads to
$Y$-functions that satisfy the 
original relation \eqref{Ysys}; this can be proven in at least two
ways. One possibility is the formal manipulation of the TBA itself,
leading to an equation of the form \eqref{slh} with the $\lambda_s(u)$
being composed of Dirac-deltas and its derivatives. These singularities appear at the isolated
points $u=0$ and are removable. The other way to prove the $Y$-system is
to construct the Quantum Transfer Matrix (QTM) associated to the
tGGE's, and then the $Y$-system follows from the fusion hierarchy,
in complete analogy with the simple thermal case
\cite{TBA-QTM-Kluemper-Takahashi}. The QTM method has been already
applied to a tGGE's including the ultra-local charges in
\cite{sajat-xxz-gge,essler-xxz-gge}. The addition of the quasi-local
charges leads to a QTM which acts on a spin chain with mixed higher
spins. However, the $Y$-system relations are not modified by this, and
they hold in the original form for all tGGE's\footnote{We are grateful
to Junji Suzuki for discussions about these questions.}. 

Quite interestingly, a generic quantum quench leads to root densities
and $Y$-functions that do not satisfy the $Y$-system.
In fact, the root densities and $Y$-functions are determined from
the charges, and generally there is no reason for the $Y$-system to hold, unless there
is some additional integrable structure associated to the particular
initial state. 
  In the case of the two-site product states
  this algebraic structure is the fusion hierarchy of the
  corresponding Boundary Quantum Transfer Matrix
  \cite{sajat-Loschm}. The atypicality of such states has been
  mentioned in \cite{jacopo-massless-1}, and it was first observed in
  \cite{eric-lorenzo-exact-solutions} that for certain four-site
  product states the Y-system is indeed violated.  

It is then a natural question to ask, how can the tGGE's approximate
the physical root densities, given that they satisfy the $Y$-system at
each truncation step. The answer is that the convergence of the
approximating root densities $\rho_s^{(n)}(u)$ is typically not uniform in the
rapidity parameter. The $Y$-system involves a shift of $\pm \eta/2$ in
the imaginary direction, which does not necessarily commute with the
infinite truncation limit. In fact, our tGGE's are constructed so that
they approximate the root density functions on the real line, because
this is the only information which is needed to obtain the correct
predictions for the local observables.
On the other hand, they can
have a different behavior in the complex plane. This is easily seen
in the examples to be considered in section \ref{examples}. The true
source terms for the TBA typically have log-zero and/or log-infinity 
singularities. It is possible to approximate such functions
using our set $D$, but the analytic properties of the approximating
functions will be markedly different.
For example there is no actual singularity at any finite
truncation step, and the log-singularities are not 
approached uniformly. Moreover, the $Y$-functions resulting from the
tGGE's are only convergent on the real axis.

\item Our construction operates in the space of the source terms for
  the TBA equations. However, it is natural to ask what is the closure
  of the linear span of the set $\{Q_{s,j}\}$ in the space of operators, and to
  which operators do the exponents in our sequences of tGGE's
  converge.

  Any finite linear combination of the $Q_{s,j}$ is an operator which is
  conserved and quasi-local, but
  it needs to be investigated whether
these properties are  preserved by our limiting procedures. We
conjecture that the commutativity with the Hamiltonian holds even for
the closure of the linear span of the set $\{Q_{s,j}\}$. On the other
hand, the quasi-locality property can be violated, which follows from
the fact that the original rapidity dependent family $X_s(u)$ is only quasi-local
inside the physical strip with  $\left|\Im u\right|< \eta/2$.

As it was already remarked above, in \cite{enej-gge} a canonical GGE
was built using the particle density operators, as given by eq. \eqref{rhorho}.
 Whereas this GGE correctly captures the
post-quench steady states, all the terms in the exponent lose their
quasi-locality as the regulator $\eps$ is sent to zero. 
It is then natural to conjecture that the limit of our sequences of tGGE's
actually coincides with the zero-regulator limit of \eqref{rhorho}. If this is indeed true,
it means that our construction is another way to approximate the
inevitably non-local exponent of the GGE using physically acceptable,
quasi-local operators. 

\item It is also useful to compare our results to the original
  formulation of the complete GGE \eqref{intGGE}, which was proven to
  be not viable in \cite{enej-gge}. Comparing the fictitious TBA
  \eqref{fictitious} to \eqref{CGGE-enej} we find that the original
  formulation can only describe the Bethe root densities if
  \begin{equation}
    \label{nemmegy}
  d\star   \lambda_s =\delta_s\,.
  \end{equation}
However, as explained in \cite{enej-gge}, the convolution with $d$
does not have an inverse, therefore \eqref{nemmegy} can not be
satisfied for an arbitrary quench problem.

Within our approach the fictitious source terms are approximated with
the function $d$ and its derivatives. At any finite truncation
\eqref{nemmegy} can be satisfied if we allow the $\lambda_s$ to become
distributions instead of analytic functions. In particular, if for a
given truncation index $n$
\begin{equation*}
  \delta_s(u)\approx \sum_{j=1}^{n} \beta_{s,j}(n) d^{(j-1)}(u)\,,
\end{equation*}
then we have
\begin{equation*}
  \lambda_s\approx \sum_{j=1}^{n} \beta_{s,j}(n)  \delta^{(j-1)},
\end{equation*}
with $\delta$ being the Dirac-delta.

As explaned above, the coefficients $\beta_{s,j}(n)$ typically do not
have a $n\to\infty$ limit, therefore
the formulation \eqref{CGGE-enej}
is ill-defined even if we allow the Lagrange-multipliers to become distributions.

\item We note an interesting consequence of our
construction: It is possible to leave out
any finite number of charges from the tGGE's, given that the first
members $Q_{s,1}$ are included for any spin index $s$. This is most
easily seen on the example of the set $D'=D\setminus
\{d'\}=\{d,d'',d''',d'''',\dots\}$. It can be proven that all source terms of the TBA
\eqref{CGGE-TBA3} can be approximated using $D'$ only: the
first member $d$ can be used to correctly fit the constant terms of
$\delta_s(u)$, and the set $D_2=\{d'',d''',d'''',\dots\}$ is
sufficient to approximate the remainder. A similar argument shows that
in fact any other finite number of charges can be left out,
as long as the first member $Q_{s,1}$ is always included.
Such GGE's with missing charges were already studied in the case of
the Ising model in
\cite{essler-truncated-gge}.
There it was found that leaving out
certain charges does have an effect on the mean values of local
correlators, which is in contradiction with our results.
This contradiction can be traced to the fact that in the free theories
the relevant set of source functions simply consists of Fourier modes,
therefore it is a Schauder basis, which is not true in the interacting case.

We also remark that the omission of a finite number of charges has an
effect on the speed of convergence of the tGGE. Leaving out one or two
charges, there are more and more operators needed to reach the full
microcanonical predictions with the same accuracy. This is
demonstrated on the example of the dimer state in section \ref{examples}.

\end{itemize}

\subsection{Identification of physical state functions}

In the previous paragraphs, we have seen that the sequence of truncated GGE's of the form \eqref{seriesdef} cannot reproduce properly the steady-state properties after a generic quantum quench, unless the Lagrange multipliers $\beta_{s,j}$ (for $j>1$) are allowed to vary with the truncation index $n$. In other terms, for $j>1$ these do not have a well-defined value in the $n \to \infty$ limit, and therefore cannot be considered as physical state functions. 
We will now show, as a corollary of our construction, that state functions can in fact be built.

 In practice the approximation of the source terms of the TBA can be
performed using an orthonormal basis of functions $\tilde D=\{\tilde
d_j\}_{j=1,\dots,\infty}$ which is obtained from $D$ by Gram-Schmidt
orthogonalization procedure with respect to the usual
$L_2$ norm :
\begin{equation*}
  \tilde d_{j}= A_{jk} d_k,\qquad \text{where}\quad A_{jk}\ne 0 \quad\text{for} \quad
  j\ge k\,,
\end{equation*}
and
\begin{equation}
\label{ortho}
  (\tilde d_j,\tilde d_k)\equiv\int_{-\pi/2}^{\pi/2} du\ \tilde
  d_j(u)\tilde d_k(u)=\delta_{jk}\,.
\end{equation}
A numerical example for the linear transformation is given below (here
$\Delta=3$):
\begin{equation*}
  A=
  \begin{pmatrix}
0.50885 &  &  &  &  & \\ 
0.21118 & 0.25771 &  &  &  & \\ 
0.10332 & 0.27570 & 0.02683 &  &  & \dots \\ 
0.05379 & 0.27368 & 0.04831 & 0.00115 &  & \\ 
0.02903 & 0.28385 & 0.06799 & 0.00293 & 0.00003 & \\
& & \vdots & & & \ddots \\
\end{pmatrix}\,.
\end{equation*}

 It follows from the completeness of $D$ and the
orthogonality condition \eqref{ortho} that $\tilde D$ is a Schauder basis, i.e. every
source term $\delta_s(u)$ in the TBA \eqref{CGGE-TBA3} has a unique expansion
\begin{equation}
\label{alphas}
  \delta_s(u)=\sum_{j=1}^\infty \tilde{\beta}_{s,j}\tilde d_j(u) \,,
\end{equation}
where the Lagrange multipliers are determined simply from
      \begin{equation*}
        \tilde \beta_{s,k}=\int_{-\pi/2}^{\pi/2} du\ \tilde d_k(u) \delta_s(u) \,.
      \end{equation*}
Resultingly, the TBA equations \eqref{CGGE-TBA3} can be rewritten as 
\begin{equation}
  \label{CGGE-TBA4}
  \begin{split}
  \log(Y_s)&=    -\delta_{s\le N_s}\sum_{k=1}^{N_d}\tilde \beta_{s,k} \tilde d_k+d\star
  \left(\log(1+Y_{s+1})+\log(1+Y_{s-1})
  \right)\,.
        \end{split}
      \end{equation}

It follows from the structure of the source terms in \eqref{CGGE-TBA3}
that    
the equivalent TBA's  \eqref{CGGE-TBA3}-\eqref{CGGE-TBA4} can be interpreted as generated by a
sequence of truncated GGE's of the form  
      \begin{equation}
  \label{eq:CGGE9}
  \varrho_n=\frac{1}{Z}\exp\left(
    -\sum_{s=1}^{N_s(n)}\sum_{j=1}^{N_d(n)} \tilde{\beta}_{s,j}\tilde Q_{s,j}
    \right) \,,
  \end{equation}
  where the new set of charges is defined using the same linear
  transformation:
  \begin{equation}
\label{tildeQdef}
    \tilde Q_{s,j}=A_{jk} Q_{s,k}\qquad s=1,2,\dots \,,
\end{equation}
and where for each truncation index $n$ we have the requirement
\begin{equation*}
\sum_{j=1}^{n} \beta_{s,j}(n)Q_{s,j}=\sum_{j=1}^{n} \tilde \beta_{s,j}\tilde Q_{s,j}   \,. 
\end{equation*}
It follows that the $\{\tilde \beta_{s,j}\}$ are related to the $\{\beta_{s,j}\}$ through 
\begin{equation}
  \label{betafromtilde}
  \beta_{s,j}(n)=\sum_{k=1}^n A_{kj} \tilde \beta_{s,k}.
\end{equation}

The new discrete set of operators $\{\tilde Q_{s,j}\}$ can serve as the basis
of the GGE in the usual sense, namely 
\begin{equation}
  \label{eq:CGGEtilde}
  \varrho=\lim_{n\to\infty }\frac{1}{Z}\exp\left(
    -\sum_{s=1}^{n}\sum_{j=1}^{n} \tilde{\beta}_{s,j}\tilde Q_{s,j}
    \right) \,,
  \end{equation}
where each Lagrange multiplier has a
well defined value and therefore plays the role of a physical state
function. For a given truncation index $n$ the $\tilde{\beta}_{s,j}$
may be written as a linear combination of all the $\beta_{s,j'}$ with
$j \leq j' \leq n$, and similarly, the ${\beta}_{s,j}$ may be written
as a linear combination of all the $\tilde{\beta}_{s,j'}$ with $j \leq
j' \leq n$. It follows from \eqref{betafromtilde} that even though the $\tilde{\beta}_{s,j}$ have
a well defined ($n$-independent) value, this is not the case for the
${\beta}_{s,j}$, and in fact they might not converge in the
$n\to\infty$ limit.

Conversely, the charges $\tilde{Q}_{s,j}$ (whose
construction does not depend on the truncation index $n$, neither on
the considered initial state) are linear combinations of the
$Q_{s,j'}$ for $j' \geq j$, so they inherit from the (quasi)locality
properties of the latter.   

We remark that even though
all quantities are well defined and finite
in \eqref{eq:CGGEtilde}, the resulting density matrix is still to be
understood as an infinite truncation limit. As a result, the TBA
calculations have to be restricted to the real line, because analytic
continuation into the complex plane typically does not commute with
the infinite truncation limit.

An important aspect of the set $\{ \tilde{Q}_{s,j} \}$ is that all
charges are required to properly reproduce the post-quench
properties. This clearly stems from the orthogonality property of the
set $\tilde{D}$, namely removing one function from this set makes
impossible to reconstruct a generic source term $\delta_{s}(\lambda)$. This contrasts with the case of the charges $\{Q_{s,j}\}$, for which we have argued in the previous paragraph that a defective sequence of tGGE's may in the generic (interacting) case still correctly reproduce the post-quench properties, albeit with a slower convergence.

To conclude this section, let us point out that while the construction
of the set $\{\tilde Q_{s,j}\}$
is independent of the considered initial state,
its choice is in fact not unique. 
 Indeed, if the Gram-Schmidt orthogonalization is performed on
a set $D'$ where a finite number of elements of $D$ are left out, we
obtain a different set of modified charges. 
In practice this means
that there are multiple ways to obtain a set of operators which serves
as a ``Schauder-like'' basis for the GGE, meaning that all coefficients
will have fixed finite values for all initial states once the
operators are fixed.
However, there remains a certain amount of ambiguity in choosing the operators
themselves.
As it was remarked above, the omission of certain elements of $D$
leads to a slower convergence of the tGGE, therefore the physically
motivated choice is to include all elements of $D$, and this selects a
particular set of modified charges.
Another possibility for constructing the set $\{\tilde Q_{s,j}\}$  is
to use a different choice for the inner product in the space of source
functions. All of these choices are expected to lead to different
sets of physical charges $\{\tilde Q_{s,j}\}$, and conversely, to
different sets of state functions $\{\tilde \beta_{s,j}\}$. However, 
all of these choices are linearly related, therefore equally physical.

\section{Examples}

\label{examples}

Here we numerically investigate 
the approximation procedure introduced in the previous section
using concrete examples.
For the quenches we consider initial states that are products of
local states spanning a few sites. The reasons for choosing such
states is twofold. First, they could be tailored in experiments,
partly due to their simple structure, or because they are ground
states of simple local Hamiltonians. The second reason for choosing
such states is that there are exact methods available to compute the
conserved charges in these states
\cite{essler-xxz-gge,JS-CGGE}, which makes their numerical treatment
relatively easy.

We aim to show that it is indeed possible to construct a sequence of
tGGE's of the form \eqref{seriesdef2} such that the local correlators
will approach their steady state values as the truncation index $n$ is increased. In section \ref{micro} it was
explained that the microcanonical GGE is correct in the XXZ model,
therefore in principle it is enough to demonstrate that the tGGE's
produce root densities that converge to the physical $\rho_{k}(u)$ as
obtained from the charges. However, for an easier comparison (and also
due to their physical importance) we will
investigate the convergence of the local correlators. 

Our concrete examples are the so-called dimer state
and two different four-site states; they will be investigated in
subsections \ref{dimer} and \ref{foursite}, respectively. The dimer
state has already been studied extensively
\cite{sajat-oTBA,sajat-QA-GGE-hosszu-cikk,jacopo-michael-hirota}, but
there are fewer results available for the four-site states.

As it was already pointed out in \cite{eric-lorenzo-exact-solutions},
in the case of four site states
the resulting $Y$-functions (denoted as $\eta$-functions in \cite{eric-lorenzo-exact-solutions}) typically do not satisfy the
$Y$-system equations \eqref{Ysys}, and this motivates their
detailed study.
In \cite{eric-lorenzo-exact-solutions} exact results were derived for
the root densities in the case of a particular four-site state (the
so-called 4-site domain wall state),
however, real time evolution in these cases has not yet been considered before.
Our goals are twofold. First we intend to demonstrate that our
construction of the truncated GGE's is independent of the $Y$-system
relations and
also works when these do not hold. Second, we aim to check the
validity of the full microcanonical GGE. Therefore 
we also perform a simulation of the real time evolution
using the iTEBD method \cite{vidal-itebd0,vidal-itebd1,vidal-itebd1a}; the details of the method have been described
in our previous works \cite{sajat-oTBA,sajat-QA-GGE-hosszu-cikk}.
Even though there is no reason to expect any
deviations from the GGE predictions, we believe it is useful to present these
numerical results as well.

\bigskip

Our methods are the following. For each initial state we construct a
series of tGGE density matrices of the form \eqref{eq:CGGE2} and
compute local correlators as a function of the series index $n$. For
the truncation numbers we choose $N_s=N_d=n$. This is a simple practical
choice, and other possibilities could be considered as well. 

For each initial state we compute the physical root and hole densities
from the mean values of the conserved charges by the relation
\eqref{stringcharge}. The generating function for the charges is
obtained using the methods of \cite{essler-xxz-gge,JS-CGGE}. Having
obtained the exact root densities we compute the source terms $\delta_s(u)$
of the fictitious TBA equations \eqref{fictitious}. Finally we build a
series of TBA equations of the form \eqref{CGGE-TBA3} such 
that the source terms approximate the true source terms $\delta_s(u)$.

First we consider the set of functions
$D=\{d(u),d'(u),d''(u),\dots\}$ and perform a Gram-Schmidt orthogonalization with respect to $L_2$
norm 
to obtain the set $\tilde D=\{\tilde d_j(u)\}_{j=1,\dots,\infty}$ described in the
previous section.
All source terms are even with respect to parity reversal,
therefore in practice it is enough to consider the set $D_e=\{d(u),d''(u),d''''(u)\dots\}$.
Then we project the functions $\delta_s(u)$ onto the space spanned by the
first $N$ elements of the set $\tilde D$:
\begin{equation*}
  \delta_s=\delta_s^{(n)}+\dots,\qquad \delta_s^{(n)}=\sum_{k=1}^n
  \tilde \beta_{s,k} \tilde d_k\,,
\end{equation*}
where the dots represent the part of $\delta_s$ belonging to the remaining
subspace of $D_e$. At each step we numerically solve the tGGE-TBA
equations with the sources $\delta_s^{(n)}$ and evaluate the local
correlation functions. These are then compared to the physical
correlators, as obtained from the physical root densities and the
long-time limit of the real time simulations.

For the numerical solution of the TBA the infinite system
\eqref{CGGE-TBA4} has to be
truncated to $N_{eq}$ equations. In our calculations we chose $N_{eq}$
big enough so that the final
numerical results do not change more than $\ordo(10^{-8})$ as we
increase $N_{eq}$ further. We observed that for higher truncation in
the tGGE (when there are accurate source terms present for higher
strings as well) $N_{eq}$ could be set to lower values. This is in
accordance with our general conjecture that the asymptotics of the TBA
for high string indices becomes irrelevant as soon as the lower nodes acquire
the correct source terms.

\subsection{The dimer state}

\label{dimer}

The dimer state is a two-site product state where each block is a
local singlet:
\begin{equation*}
  \ket{D}=\otimes_{j=1}^{L/2} \frac{\ket{\uparrow\downarrow}-\ket{\downarrow\uparrow}}{\sqrt{2}}\,.
\end{equation*}
In this work we will use the zero-momentum projection, which is defined
as
\begin{equation*}
  \ket{D_0}=\frac{1}{\sqrt{2}}   (\ket{D}+U\ket{D})\,,
\end{equation*}
where $U$ is the one-site translation operator.

This state has been studied extensively. Its overlaps with Bethe states
were already calculated in \cite{sajat-neel}, mean values of the
ultra-local conserved charges were calculated in
\cite{fagotti-collura-essler-calabrese}, and the so-called Quench
Action solution \cite{quench-action} was given in
\cite{sajat-oTBA}. In \cite{sajat-QA-GGE-hosszu-cikk} exact results
were presented for the $Y$-functions using the methods of \cite{js-hosszu-kvencs}.
Furthermore, a complete
analytic solution of the corresponding fusion hierarchy was given for
the isotropic point $\Delta=1$ in \cite{jacopo-michael-hirota}.
We note that for the dimer state the $Y$-functions satisfy the
$Y$-system relations \eqref{Ysys}. In the works cited above this was
assumed and also checked for the first few functions, whereas a general
proof was later given in \cite{sajat-Loschm}. 

The source terms for the physical TBA are given explicitly by
\begin{equation}
  \delta_s=- \log \frac{\vartheta_4^2}{\vartheta_1^2}+(-1)^s
  \log \frac{\vartheta_2^2}{\vartheta_3^2},
 \end{equation}
where $\vartheta_{1,2,3,4}$ are Jacobi-functions with nome $e^{-2\eta}$
\cite{js-hosszu-kvencs,sajat-QA-GGE-hosszu-cikk}.
As there are only two different source terms
(for odd and even nodes),
the approximation procedure for the sources can be done at the
beginning only once, for $\delta_1$ and $\delta_2$.
 
\begin{figure}
  \centering
\begin{subfigure}[b]{0.3\textwidth}
\centering
\includegraphics[scale=0.35]{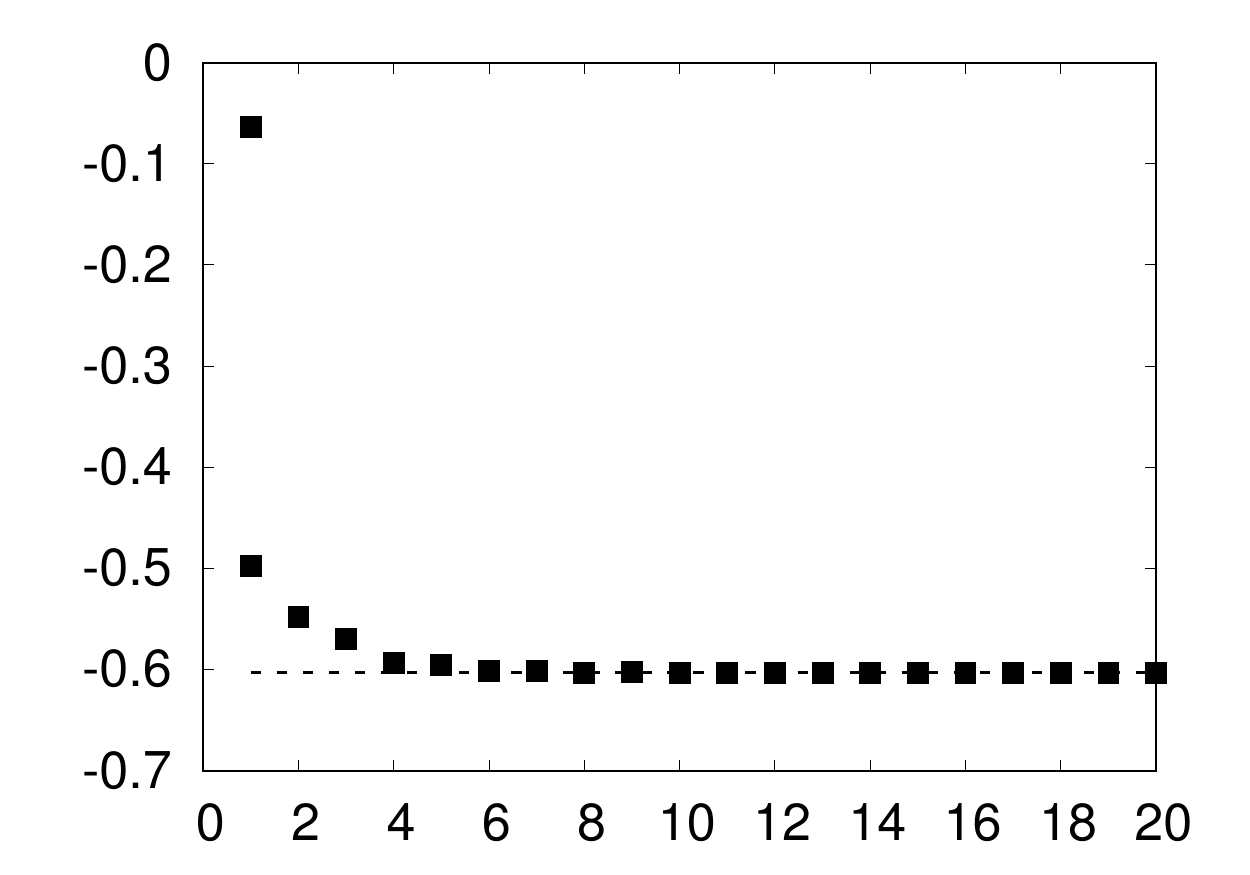}
\caption{$\sigma^z_1\sigma^z_2$}
\end{subfigure}
\begin{subfigure}[b]{0.3\textwidth}
\centering
\includegraphics[scale=0.35]{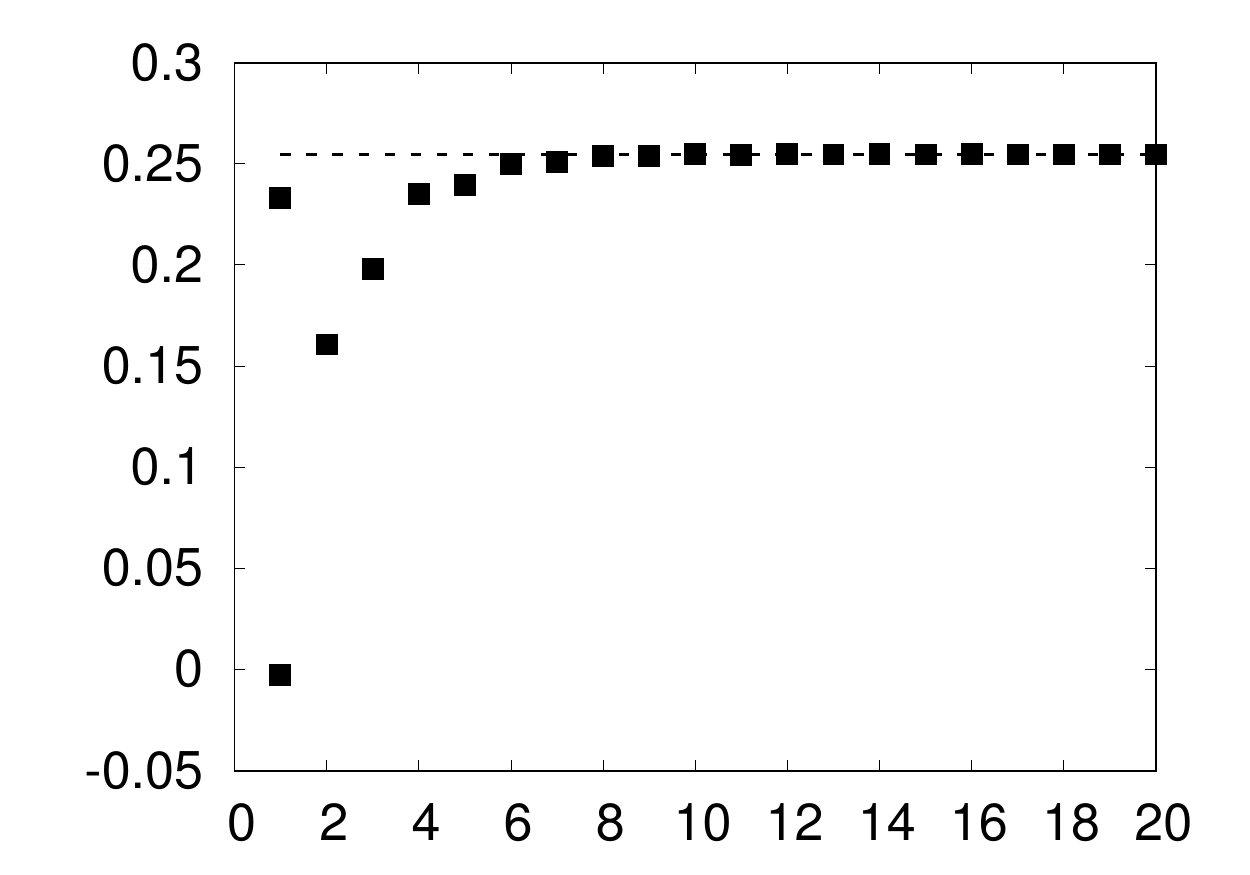}
\caption{$\sigma^z_1\sigma^z_3$}
\end{subfigure}
\begin{subfigure}[b]{0.3\textwidth}
\centering
\includegraphics[scale=0.35]{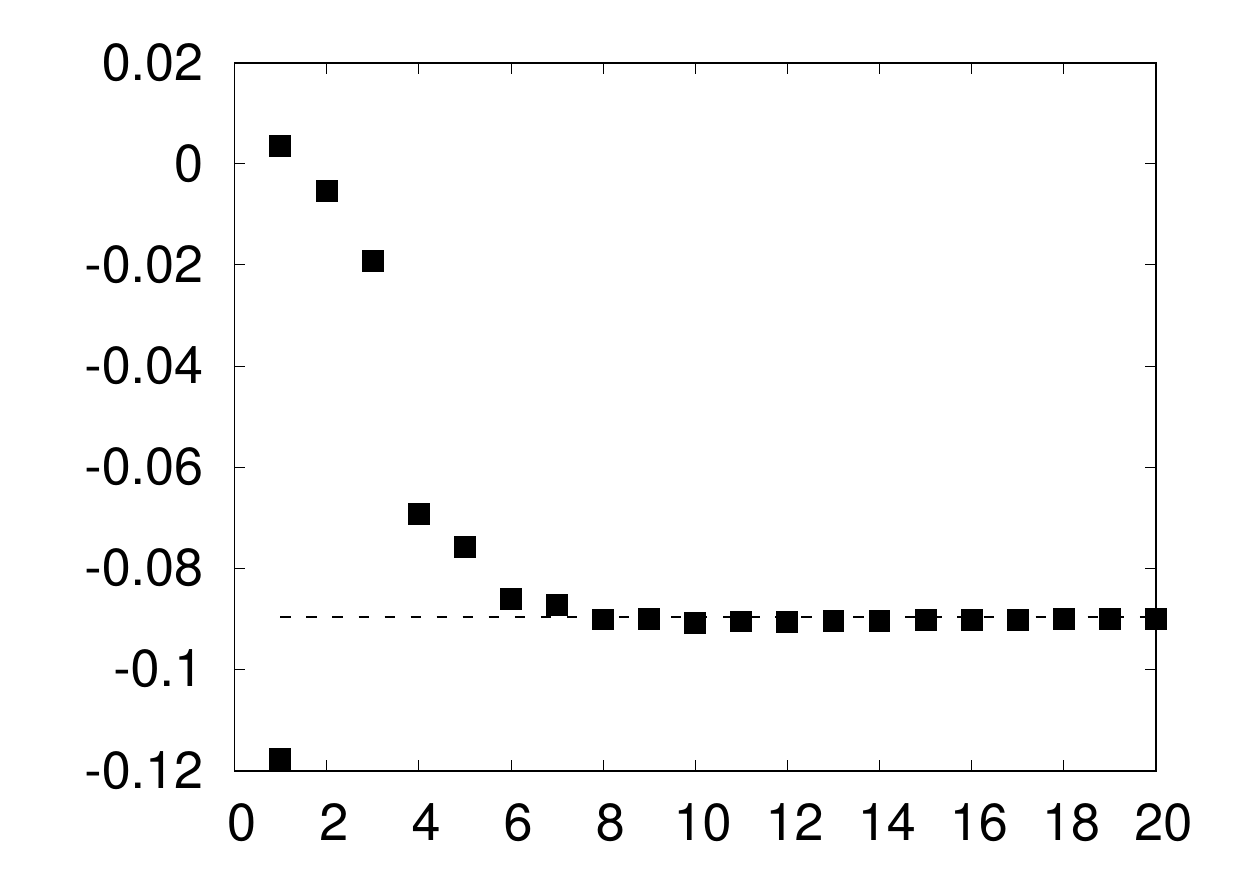}
\caption{$\sigma^z_1\sigma^z_4$}
\end{subfigure}
  \caption{Evaluation of a few local observables within the truncated
    GGE for  the dimer quench. The correlators $\sigma^z_1\sigma^z_a$,
    $a=2,3,4$ are plotted as a function of the truncation index
    $n$. At each truncation step a total of $N_sN_d=n^2$ charges are
    included in the tGGE.
The horizontal line shows the correlators computed using the Quench
Action solution, which were were shown to coincide with the
asymptotics of the real time evolution \cite{sajat-oTBA}.
    The value of the anisotropy is $\Delta=3$.}
  \label{fig:corrDimer1}
\end{figure}

Examples for the local correlators as a function of the truncation index $n$
are shown in Fig. \ref{fig:corrDimer1}. It can be seen that the
physical values are obtained with big precision already around $n=8$,
which corresponds to adding $N_sN_d=n^2=64$  charges to the the
truncated GGE.

\begin{figure}
  \centering
  \begin{subfigure}[b]{0.45\textwidth}
    \centering
    \includegraphics[scale=0.45]{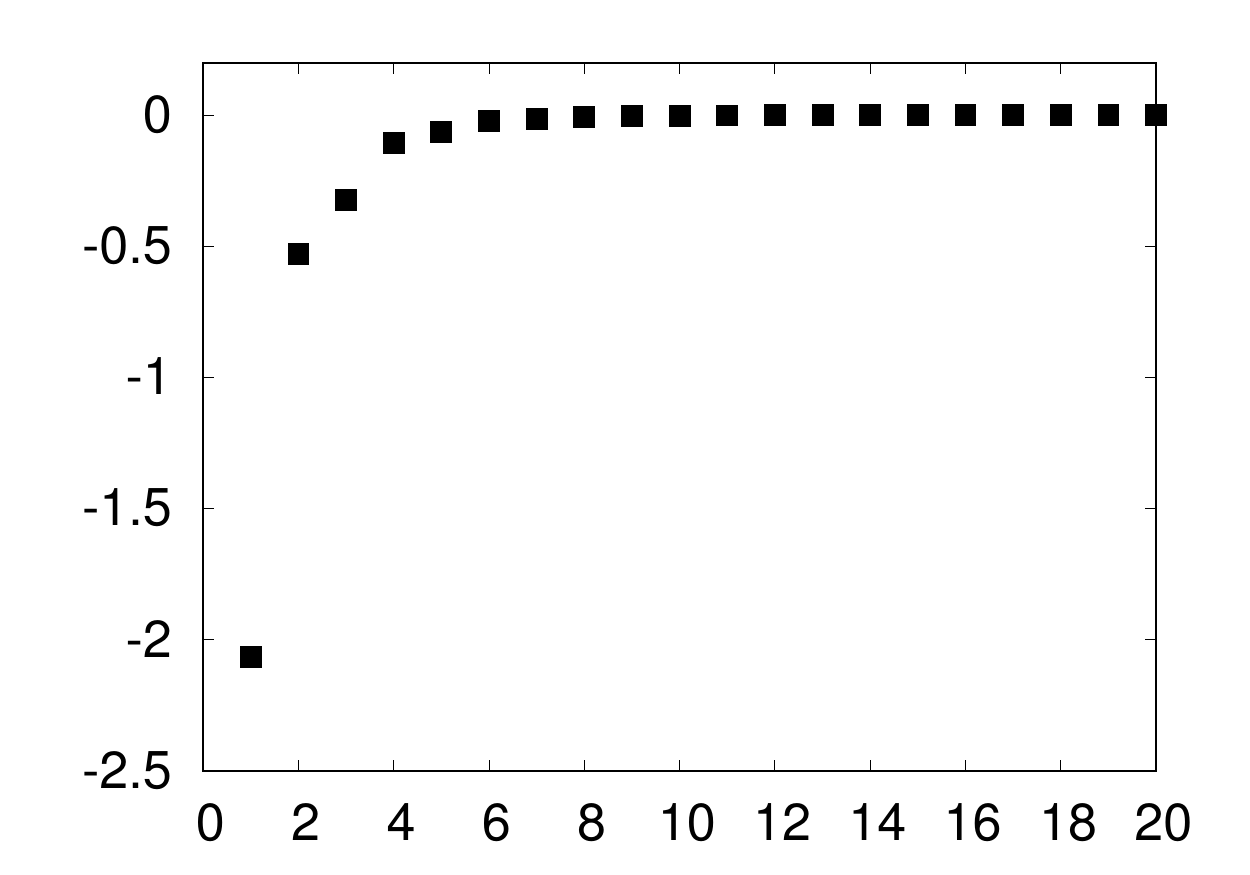}
    \caption{$\beta_{1,1}$}
  \end{subfigure}
    \begin{subfigure}[b]{0.45\textwidth}
    \centering
    \includegraphics[scale=0.45]{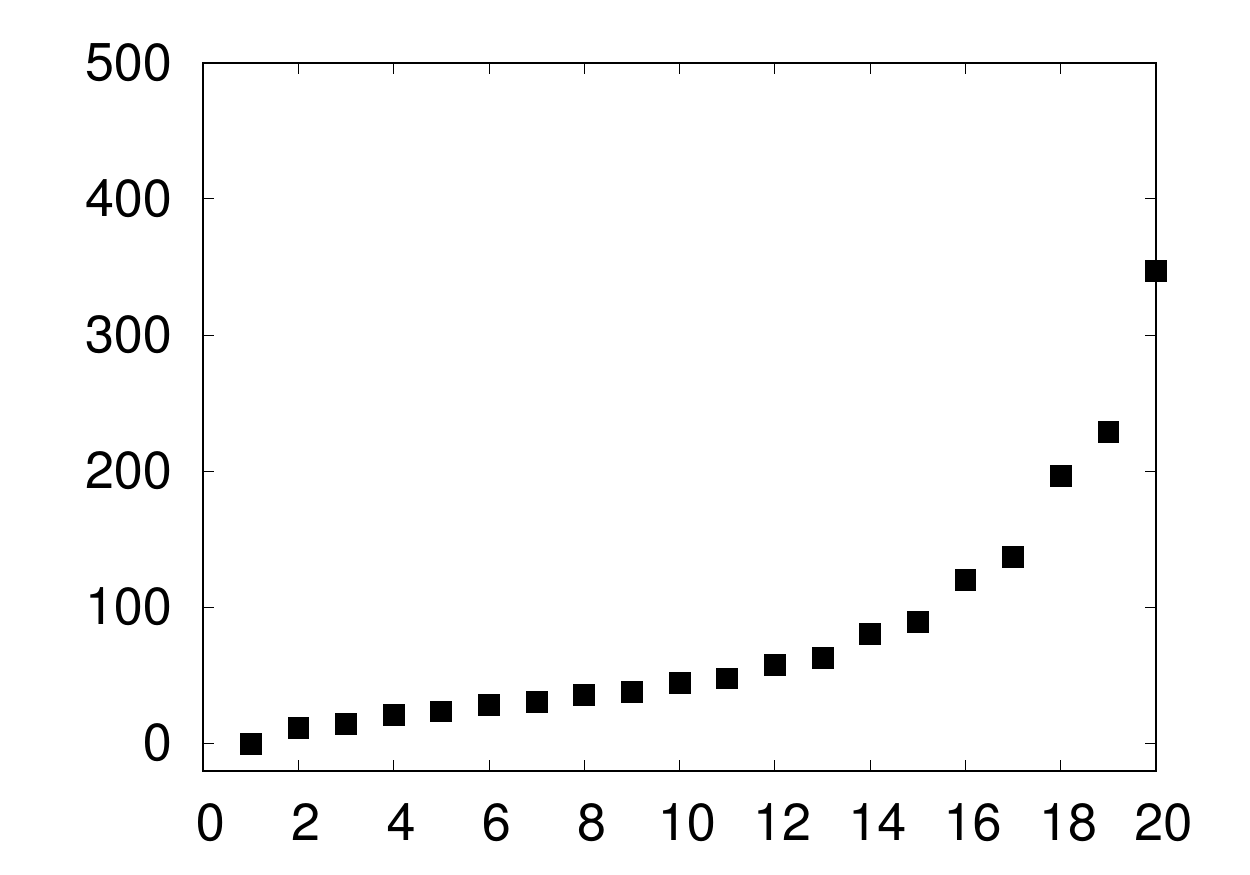}
    \caption{$\beta_{1,2}$}
  \end{subfigure}
  
    \begin{subfigure}[b]{0.45\textwidth}
    \centering
    \includegraphics[scale=0.45]{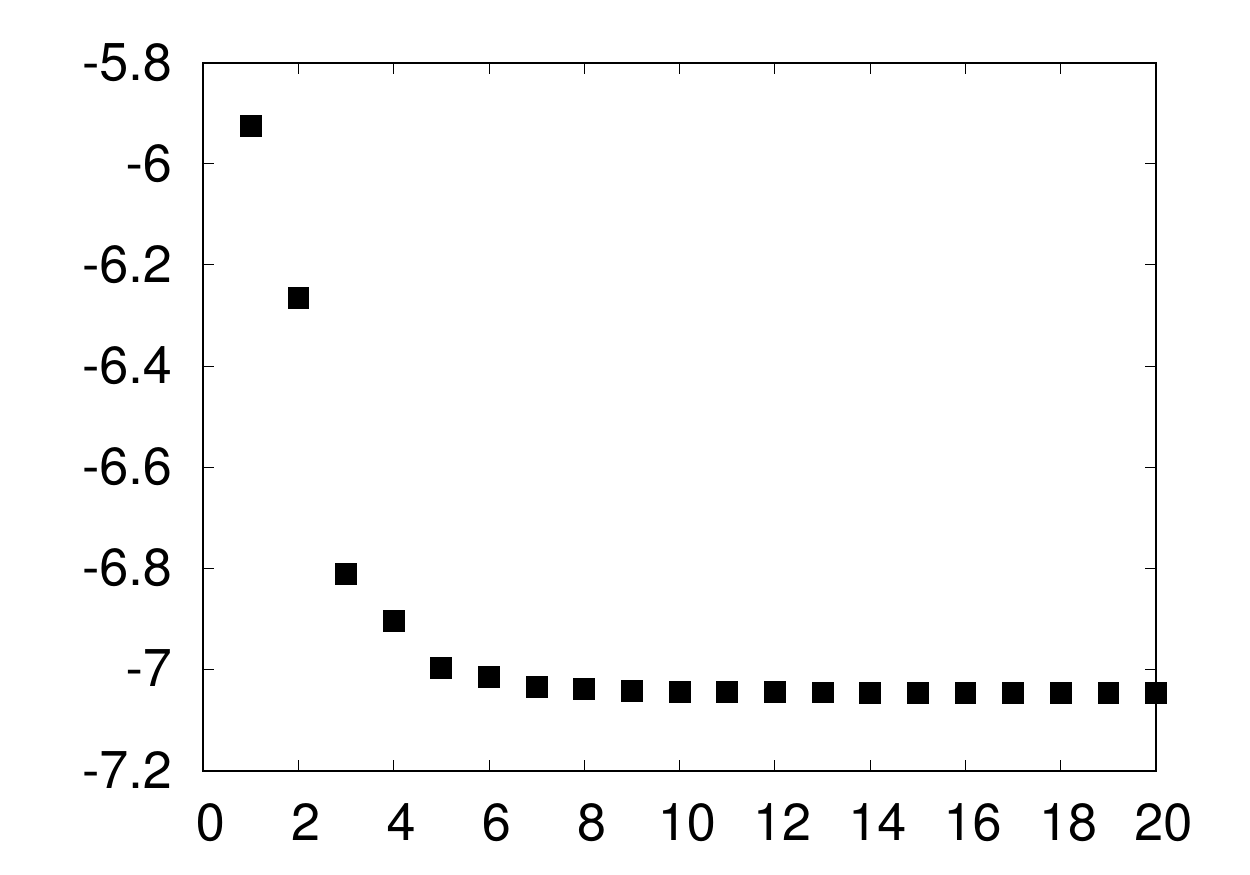}
    \caption{$\beta_{2,1}$}
  \end{subfigure}
    \begin{subfigure}[b]{0.45\textwidth}
    \centering
    \includegraphics[scale=0.45]{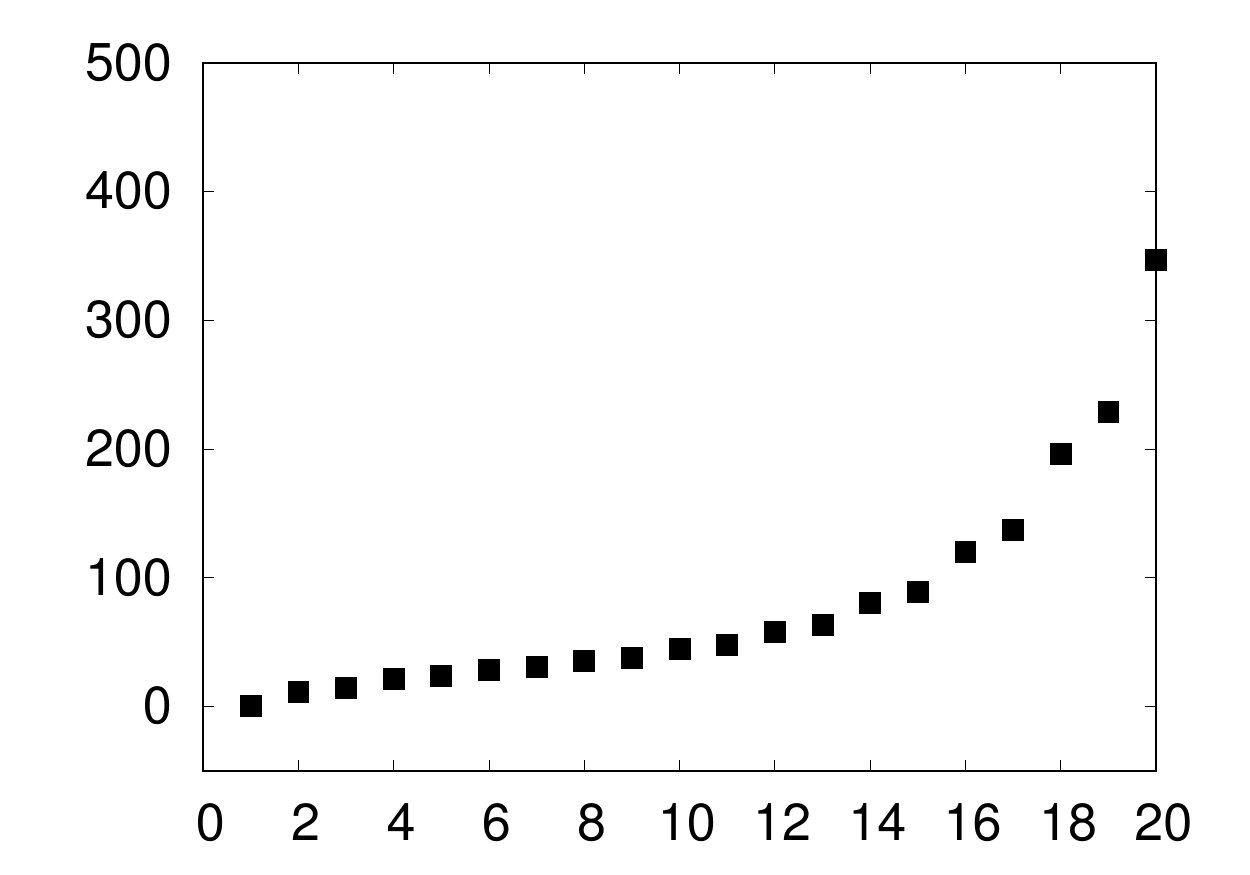}
    \caption{$\beta_{2,2}$}
    \end{subfigure}
  \caption{The dependence of the first few Lagrange multipliers $\beta_{s,j}$ on the
    truncation index $n$. For each spin index the first member
    $\beta_{s,1}$ tends to a constant value, while the remaining
    $\beta$'s are typically divergent.}
  \label{fig:beta1}
\end{figure}

The approximation procedure gives us the Lagrange multipliers $\tilde
\beta_{s,j}$ associated to the new set of charges. On the other hand,
it is also useful to calculate the original $n$-dependent set
$\beta_{s,j}(n)$ through relation \eqref{betafromtilde}.
In Fig. \ref{fig:beta1} we plot the dependence of the first few
Lagrange multipliers on the truncation index $n$. It can be seen that
$\beta_{1,1}$ and $\beta_{2,1}$ (the first members of the $s=1$ and
$s=2$ series) indeed tend to some well-defined values, as it was already
explained in the previous section. However, $\beta_{1,2}$ and
$\beta_{2,2}$ (which are the coefficients of the next charges in the
respective series) are divergent. Similar behavior is observed for
all $\beta_{s,j}$ with $j>1$, therefore we did not plot further examples.

\begin{figure}
  \centering
  \begin{subfigure}[b]{0.49\textwidth}
    \centering
    \includegraphics[scale=0.6]{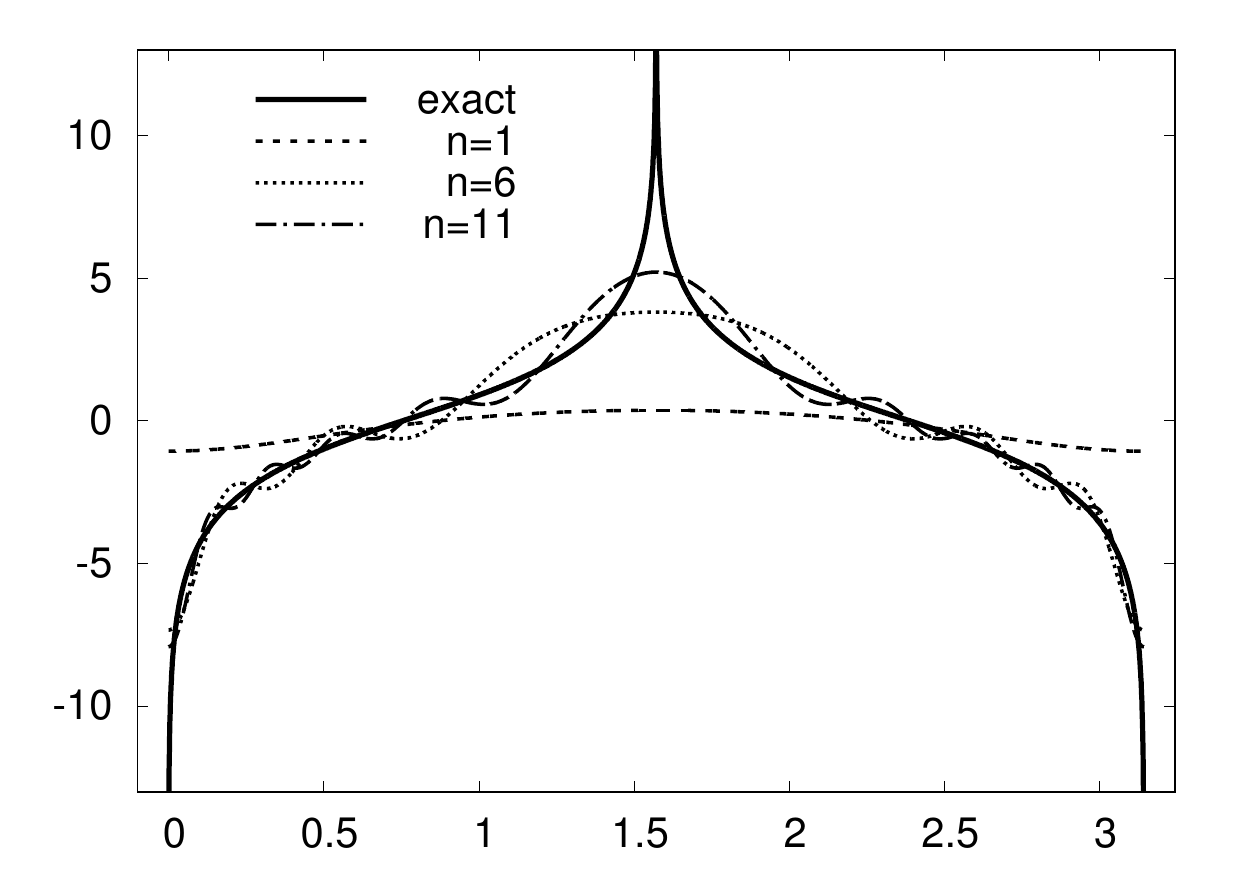}
    \caption{}
    \end{subfigure}
  \begin{subfigure}[b]{0.49\textwidth}
    \centering
    \includegraphics[scale=0.6]{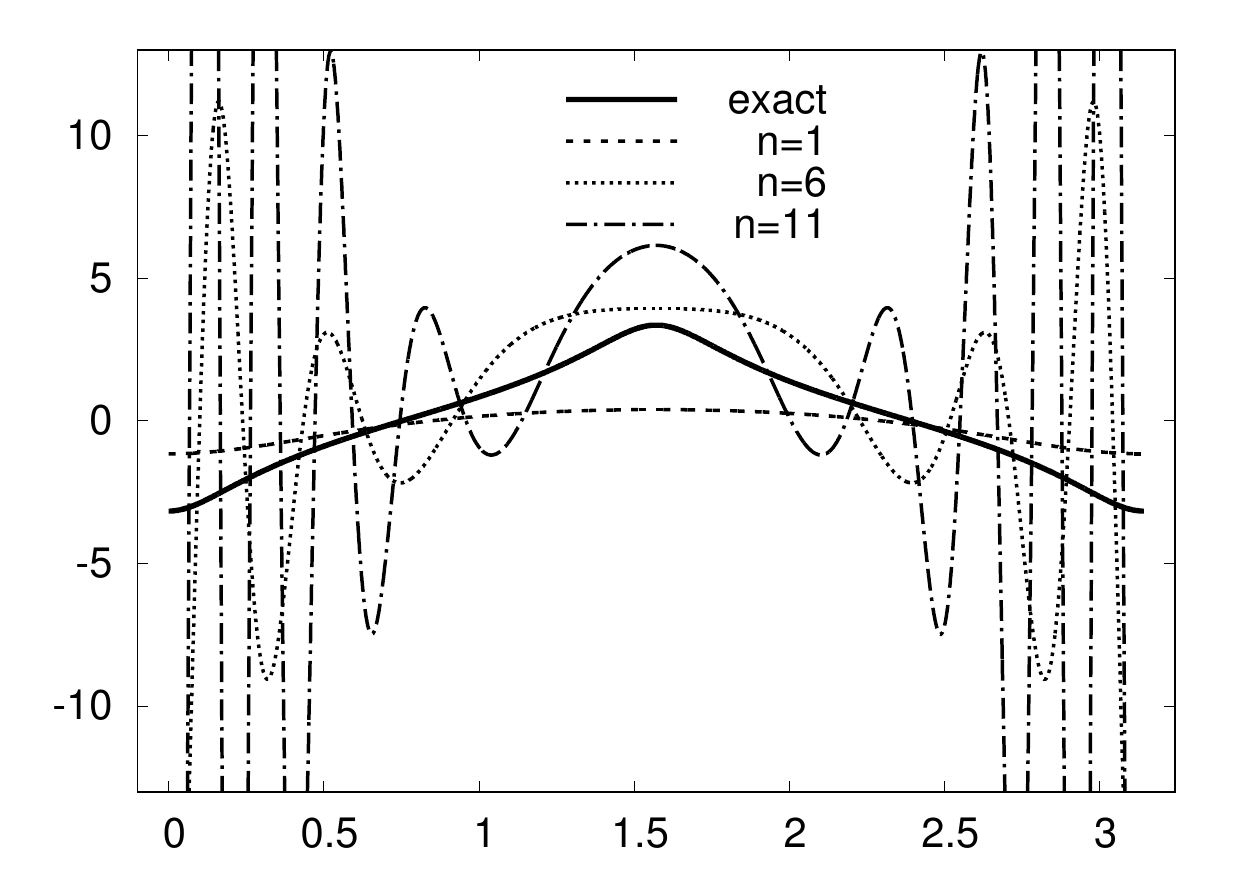}
    \caption{}
    \end{subfigure}
    \caption{The function $\log(Y_1(u))$ of the dimer quench evaluated in the truncated GGE's
      together with the exact result. Fig. (a) shows the function for
      real $u$, whereas Fig. (b) is a plot of $\log(|Y_1(u)|)$ for
      $u=x+0.1i\eta$ with $x\in\valos$. The anisotropy is $\Delta=3$.}
  \label{fig:fittingeta1}
\end{figure}

Fig. \ref{fig:fittingeta1} shows how the approximation procedure works
for the function $Y_1(u)$: the exact physical solution \cite{sajat-QA-GGE-hosszu-cikk}
\begin{equation*}
Y_1(u)=\frac{\cos(4\lambda)-\cosh(2\eta)}{\cos^2(\lambda)(\cos(2\lambda)-\cosh(\eta))}-1
\end{equation*}
is compared to the solutions of the approximating TBA equations for
the truncation numbers $n=1,6,11$. The first figure shows $\log(Y_1(u))$
evaluated on the real line: indeed it can be observed that adding more
and more charges gives a better approximation. However, the $\log(0)$
and $\log(\infty)$ singularities are never reproduced by the
approximations, which are regular for each $n$. In contrast to this
situation, the second figure shows the same function evaluated
slightly off the real axis, for $u=x+0.1 i\eta$, where
$x\in\valos$. It can be seen that convergence is lost, and indeed our
approximation procedure only works on the real line.

\begin{figure}
  \centering
\begin{subfigure}[b]{0.3\textwidth}
\centering
\includegraphics[scale=0.35]{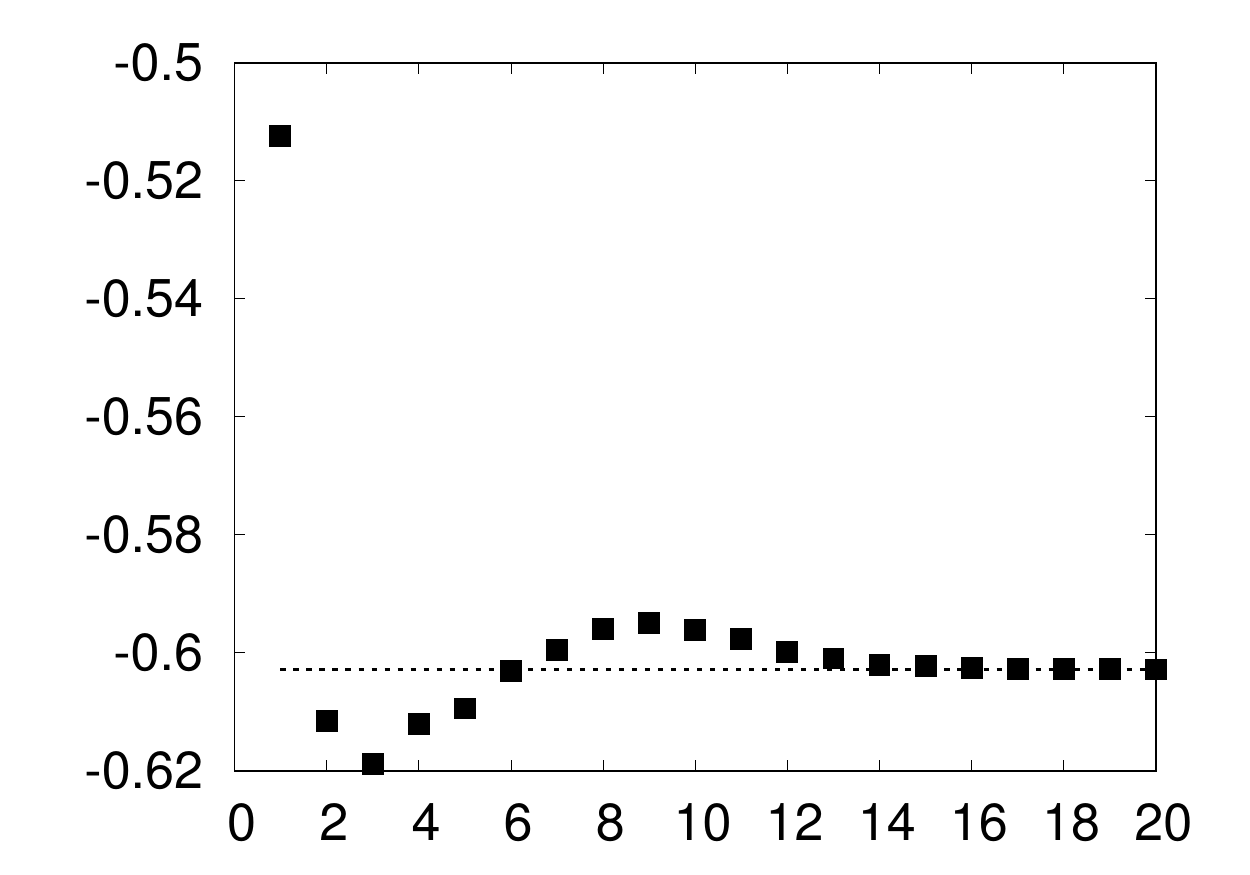}
\caption{$\sigma^z_1\sigma^z_2$}
\end{subfigure}
\begin{subfigure}[b]{0.3\textwidth}
\centering
\includegraphics[scale=0.35]{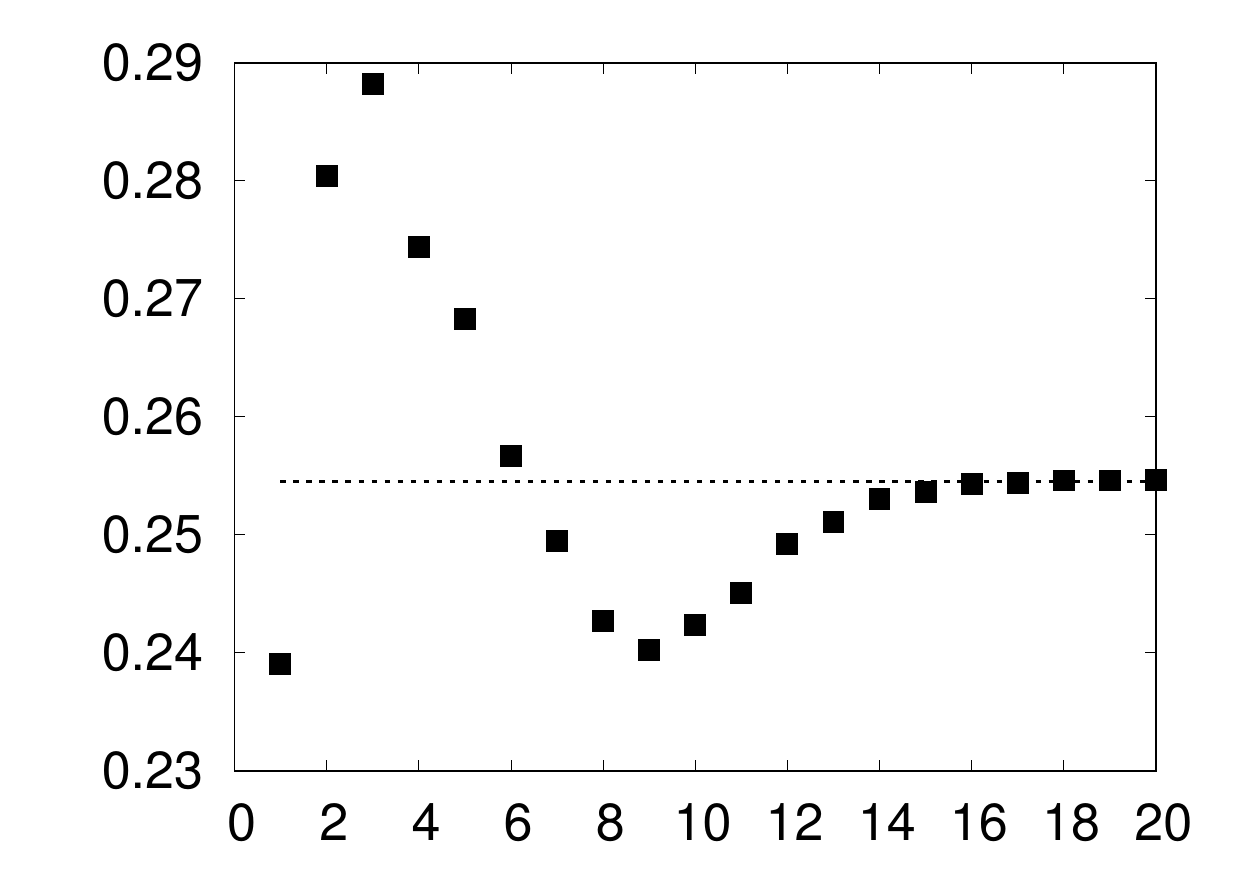}
\caption{$\sigma^z_1\sigma^z_3$}
\end{subfigure}
\begin{subfigure}[b]{0.3\textwidth}
\centering
\includegraphics[scale=0.35]{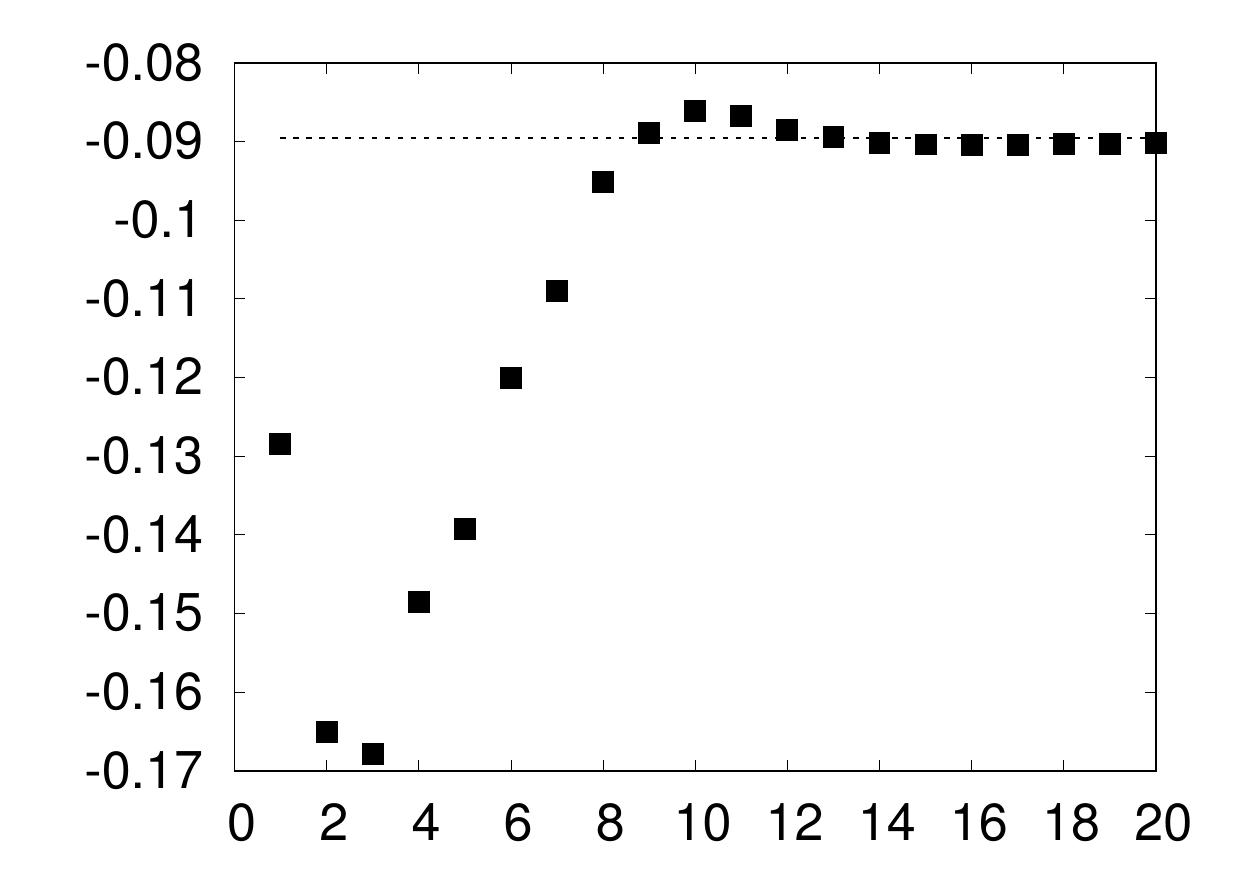}
\caption{$\sigma^z_1\sigma^z_4$}
\end{subfigure}
  \caption{The $\sigma^z_1\sigma^z_a$ correlators within a defective tGGE with $a=2,3,4$ for
    the dimer quench. Here $D=3$. The defective GGE is built with the
    omission of the operators $Q_{s,2}$, which are the second charges
    for each spin/string index. Comparing with
    Fig. \ref{fig:corrDimer1} it can be seen that the convergence in
    the truncation index is
    slower, but the physical correlators are reproduced nevertheless.}
  \label{fig:corrDimerDef1}
\end{figure}

In the previous section it was remarked that the set of charges
$\{Q_{s,j}\}$ is over-complete in the sense that a finite number of
them can be omitted from the tGGE's and yet the correct steady states
can be reproduced. The only technical requirement is that the first
members with $j=1$ have to be included for each $s$. Here we
numerically demonstrate this statement in the case of the dimer
state. We build a defective tGGE where we leave out the second even
charge for each string index. In terms of the TBA this corresponds to
performing a Gram-Schmidt orthogonalization on the set
$D'=D_e\setminus\{d''(u)\}=\{d(u),d^{(4)}(u),d^{(6)}(u),\dots\}$ and to use this basis to
approximate the actual source terms. Numerical results for the
correlators are depicted in Fig. \ref{fig:corrDimerDef1}. It can be
seen that the correct values of the
correlators are indeed reproduced, but the convergence is slower than
in the case of the full tGGE's shown in Fig. \ref{fig:corrDimer1}.

\subsection{Four-site states}

\label{foursite}

Here we consider two different four site states as initial states.
The
first one is the domain wall state
\begin{equation*}
  \ket{DW_4}\equiv \otimes_{j=1}^{L/4} \ket{\uparrow\uparrow\downarrow\downarrow}\,.
\end{equation*}
This state has been already considered in
\cite{eric-lorenzo-exact-solutions}, and exact expressions were
derived for the root densities using the complete set of
charges. Moreover, it was found that the resulting $Y$-functions do
not satisfy the $Y$-system equations. Whereas it was already argued in
\cite{enej-gge} that this is the typical behavior for a quantum
quench, real time evolution in such cases have not yet been considered
in the literature.

We note that the domain wall state is similar to the N\'eel sate in
the sense that in the $\Delta\to\infty$ limit it becomes an exact
eigenstate of the XXZ model. Therefore, it is expected that a quench
from this initial state is a ''small quench'' for 
large and intermediate $\Delta$, similar to what was observed for the
N\'eel state in \cite{JS-oTBA,sajat-oTBA}.

In order to treat a four-site state with a certain degree of local entanglement we
consider the initial state
\begin{equation*}
  \begin{split}
  \ket{D_4}&=
\otimes_{j=1}^{L/4}
\frac{\ket{\uparrow\uparrow\downarrow\downarrow}+\ket{\downarrow\downarrow\uparrow\uparrow}-\ket{\uparrow\downarrow\downarrow\uparrow}-\ket{\downarrow\uparrow\uparrow\downarrow}}{2}\,.
  \end{split}
\end{equation*}
It can be checked that in each four-site block the even and the
odd sites are entangled with each other into an $SU(2)$ singlet. 
It follows that this state is $SU(2)$ invariant and it can be
considered as resulting from a permutation of sites performed on the
usual dimer state. Also, $\ket{D_4}$ is 
one of the ground states of the local Hamiltonian
\begin{equation*}
  H=J\sum_{j=1}^{L/4} 
(S_{4j}\cdot S_{4j+1}+S_{4j+1}\cdot S_{4j+3}+S_{4j+2}\cdot S_{4j+3}+S_{4j+2}\cdot S_{4j+4})
+
\frac{J}{2}\sum_{j=1}^{L/2}  (S_{2j}\cdot S_{2j+3}+S_{2j+1}\cdot S_{2j+2})\,.
\end{equation*}
This can be seen by performing a permutation of sites on the
Majumdar-Ghosh Hamiltonian, for which the usual dimer state is an
exact ground state.

In the case of the DW state the two-point correlators are two-site
shift invariant, and for the generalized dimer there is only a
four-site shift invariance. The GGE
can only describe the zero momentum sector of the correlators. In
order to compare the iTEBD data to the GGE we construct the averaged
two-point functions
\begin{equation*}
  \vev{\overline{\sigma^\alpha_i\sigma^\alpha_{i+a}}}\equiv
  \sum_{j=1}^4
  \vev{{\sigma^\alpha_j\sigma^\alpha_{j+a}}},\qquad
  a=1,2,\dots,\quad \alpha=x,y,z\,.
\end{equation*}

\begin{figure}[ht]
  \centering
  \begin{subfigure}[b]{0.3\textwidth}
      \centering
    \includegraphics[scale=0.35]{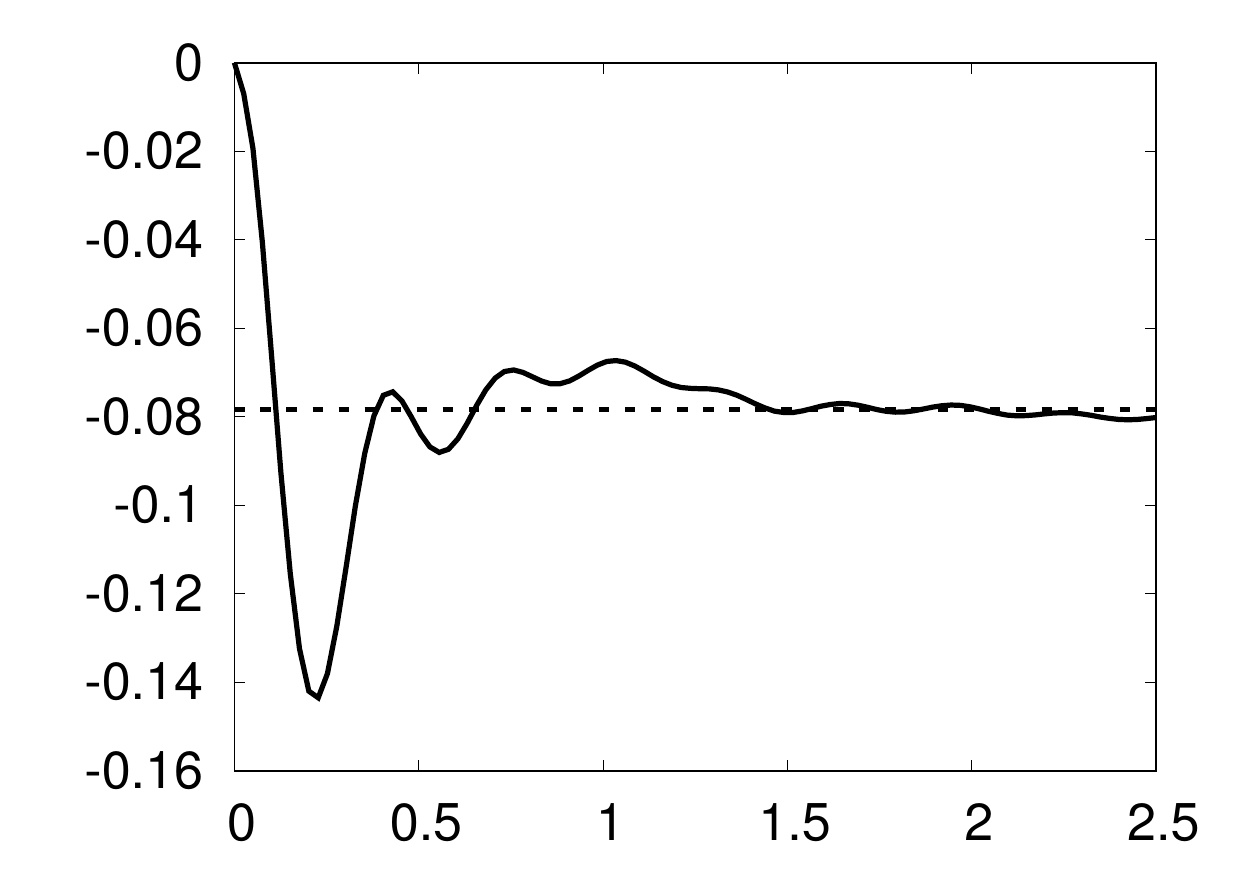}
    \caption{$\vev{\overline{\sigma^z_i\sigma^z_{i+1}}}$}
  \end{subfigure}
  \begin{subfigure}[b]{0.3\textwidth}
      \centering
    \includegraphics[scale=0.35]{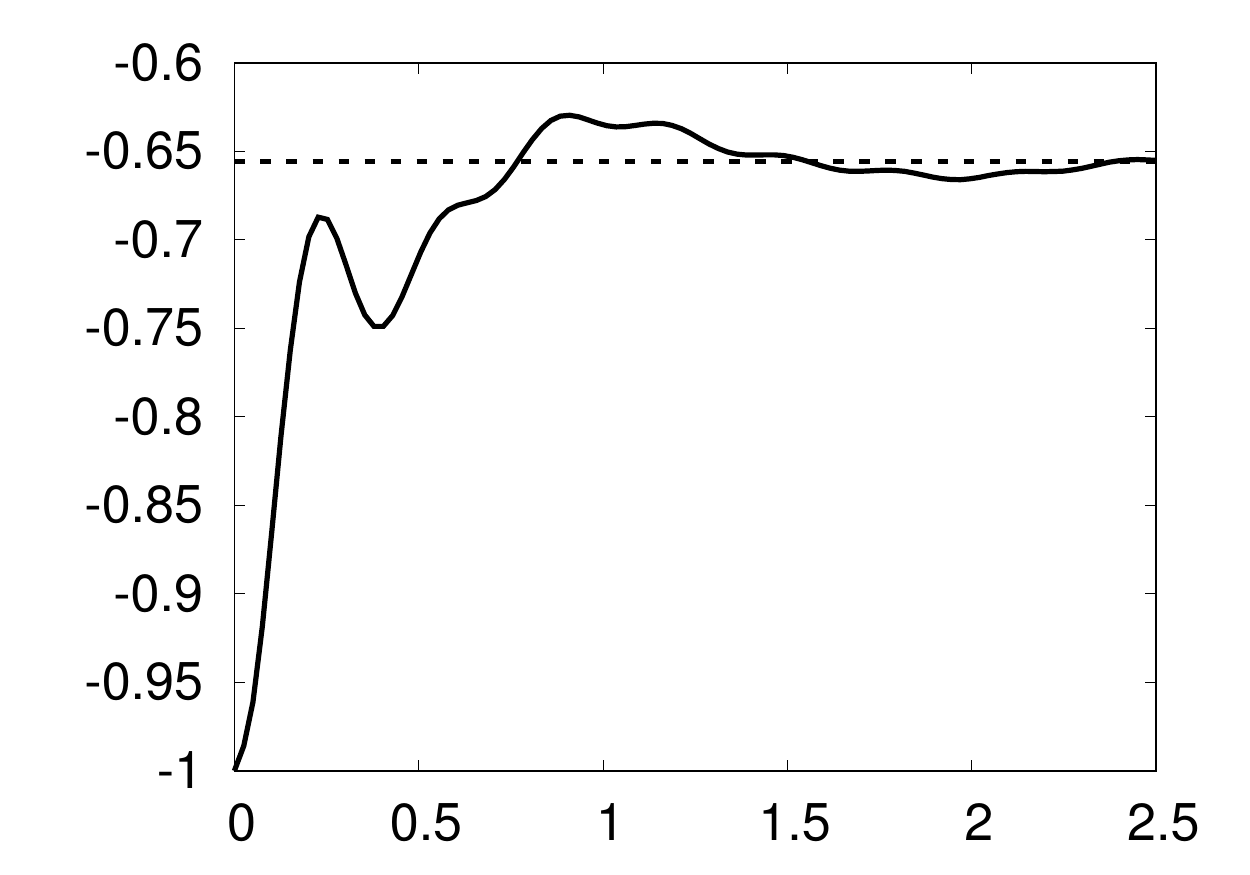}
    \caption{$\vev{\overline{\sigma^z_i\sigma^z_{i+2}}}$}
  \end{subfigure}
  \begin{subfigure}[b]{0.3\textwidth}
      \centering
    \includegraphics[scale=0.35]{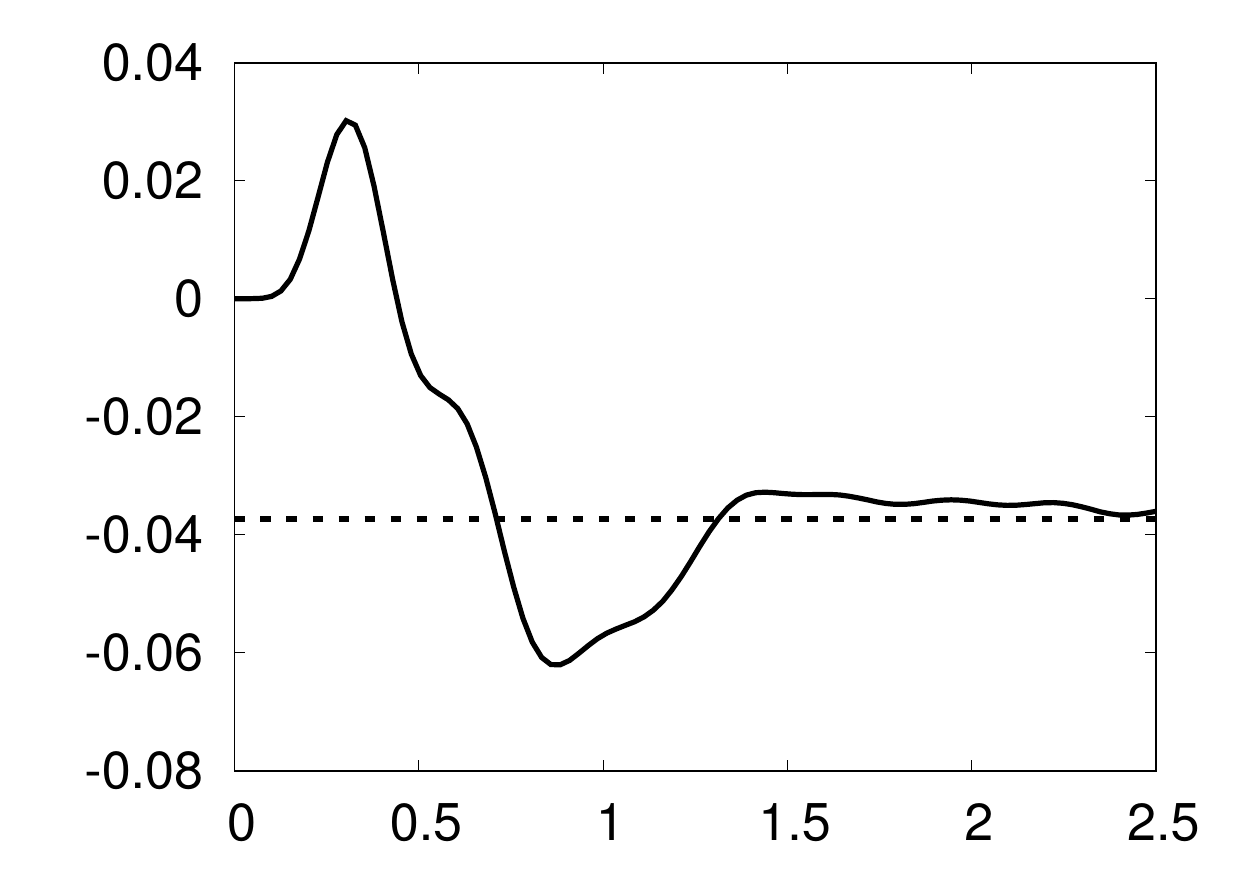}
    \caption{$\vev{\overline{\sigma^z_i\sigma^z_{i+3}}}$}
  \end{subfigure}
  
  \caption{Time evolution of the averaged short range correlators in
    the quench starting from the 4-site DW state. The horizontal line
    is the prediction of the microcanonical GGE.}
  \label{fig:DWiTEBD}
\end{figure}

\begin{figure}[ht]
  \centering
  \begin{subfigure}[b]{0.3\textwidth}
      \centering
    \includegraphics[scale=0.35]{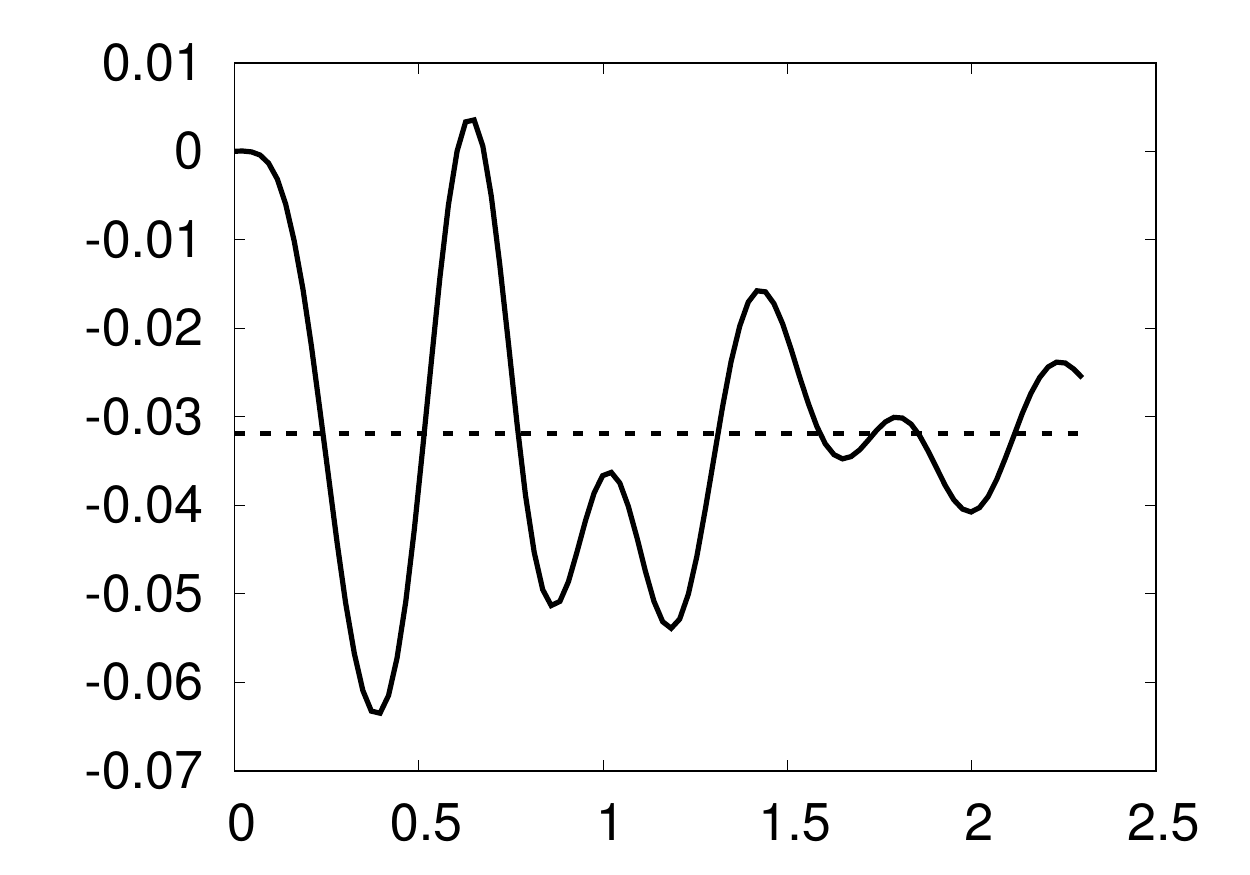}
    \caption{$\vev{\overline{\sigma^z_i\sigma^z_{i+1}}}$}
  \end{subfigure}
  \begin{subfigure}[b]{0.3\textwidth}
      \centering
    \includegraphics[scale=0.35]{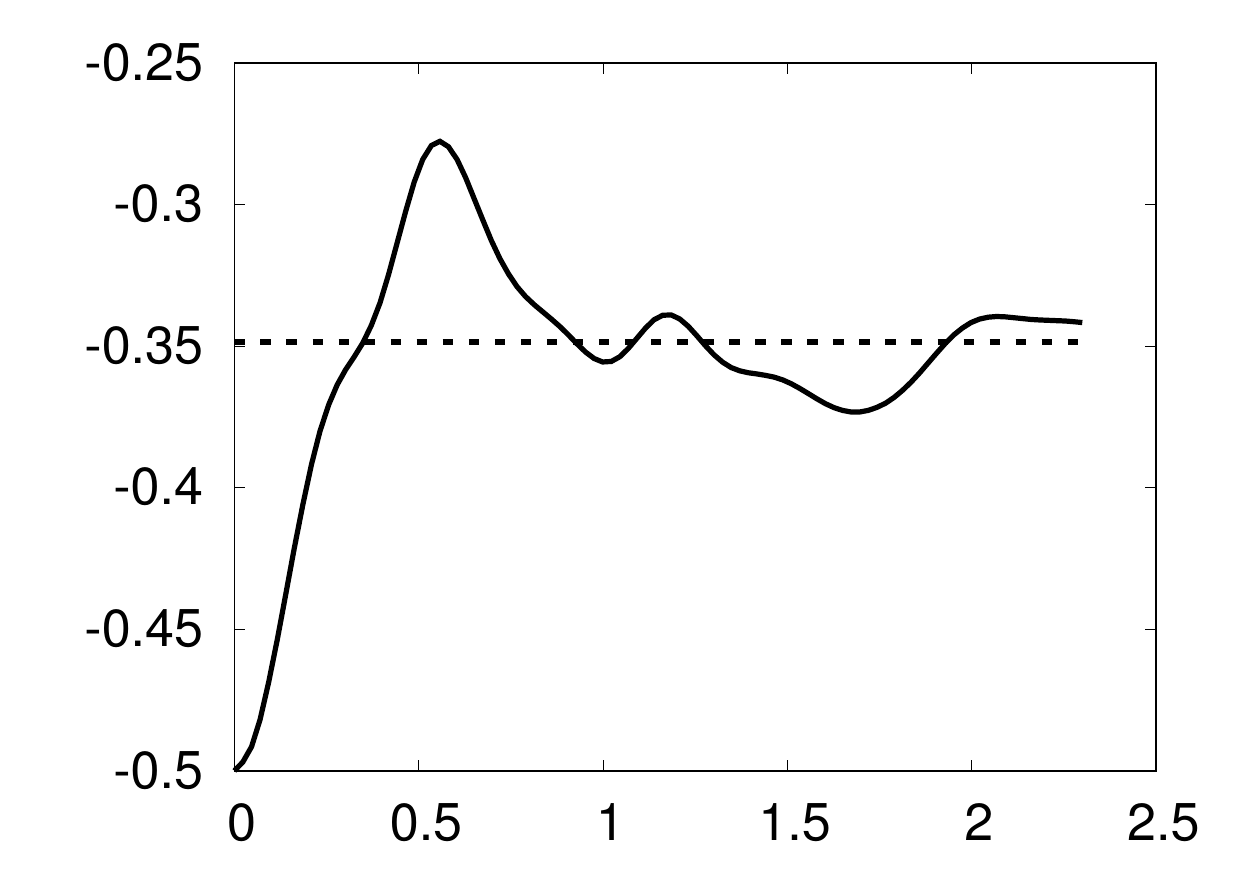}
    \caption{$\vev{\overline{\sigma^z_i\sigma^z_{i+2}}}$}
  \end{subfigure}
  \begin{subfigure}[b]{0.3\textwidth}
      \centering
    \includegraphics[scale=0.35]{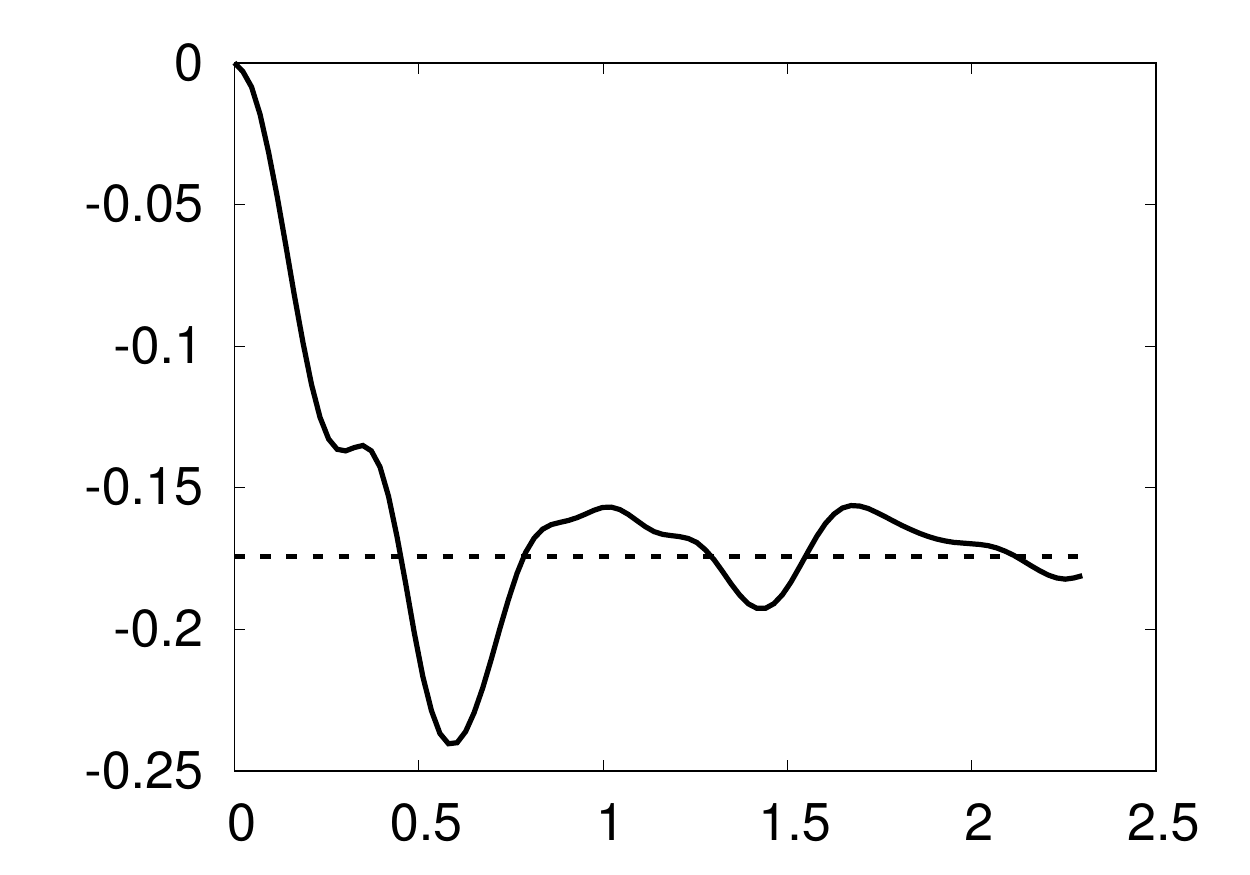}
    \caption{$\vev{\overline{\sigma^z_i\sigma^z_{i+3}}}$}
  \end{subfigure}
  
  \caption{Time evolution of the averaged short range correlators in
    the quench starting from the generalized 4-site dimer state. The horizontal line
    is the prediction of the microcanonical GGE.}
  \label{fig:iTEBD3}
\end{figure}

In Figs. \ref{fig:DWiTEBD} and \ref{fig:iTEBD3} we plot the time
evolution of short-range averaged z-z correlators against the
prediction of the full micro-canonical ensemble. In the DW case we observe a
relatively fast convergence, in accordance with this situation being a
``small quench''. On the other hand, in the generalized dimer case the
equilibration is considerably slower. It can be seen that the data
does not contradict the GGE, but he time scales
available to the iTEBD algorithm are too small to really confirm the
predictions in this case. We note that plotting the individual
correlators instead of the averaged ones we observe an even slower
equilibration (even in the DW case), and our data is not sufficient to
determine whether translational invariance is restored in the long
time limit. We remind the reader that this is an open issue which to
our best knowledge has not been resolved even for the simple two-site
states \cite{fagotti-collura-essler-calabrese,sajat-oTBA,sajat-QA-GGE-hosszu-cikk}.

\begin{figure}
  \centering
\begin{subfigure}[b]{0.3\textwidth}
\centering
\includegraphics[scale=0.35]{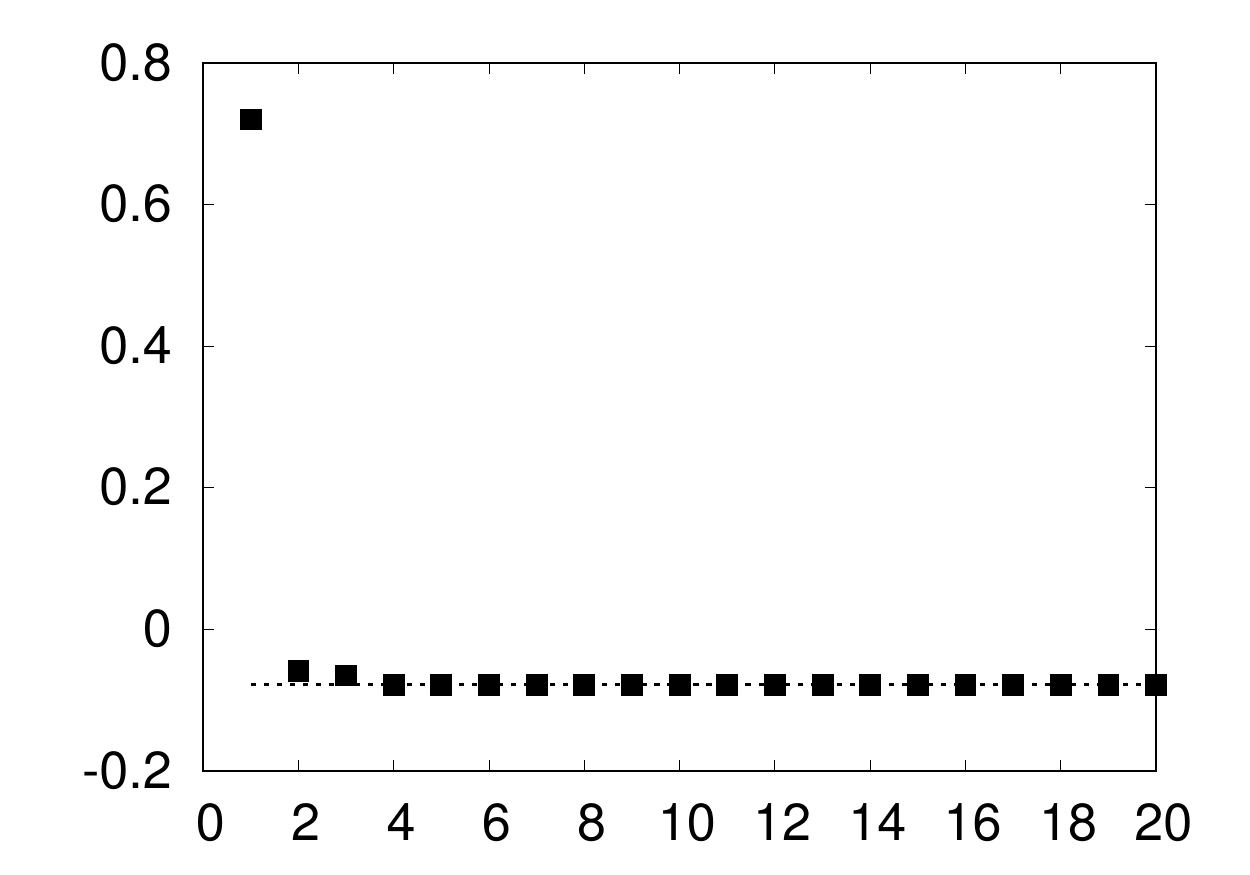}
\caption{$\sigma^z_1\sigma^z_2$}
\end{subfigure}
\begin{subfigure}[b]{0.3\textwidth}
\centering
\includegraphics[scale=0.35]{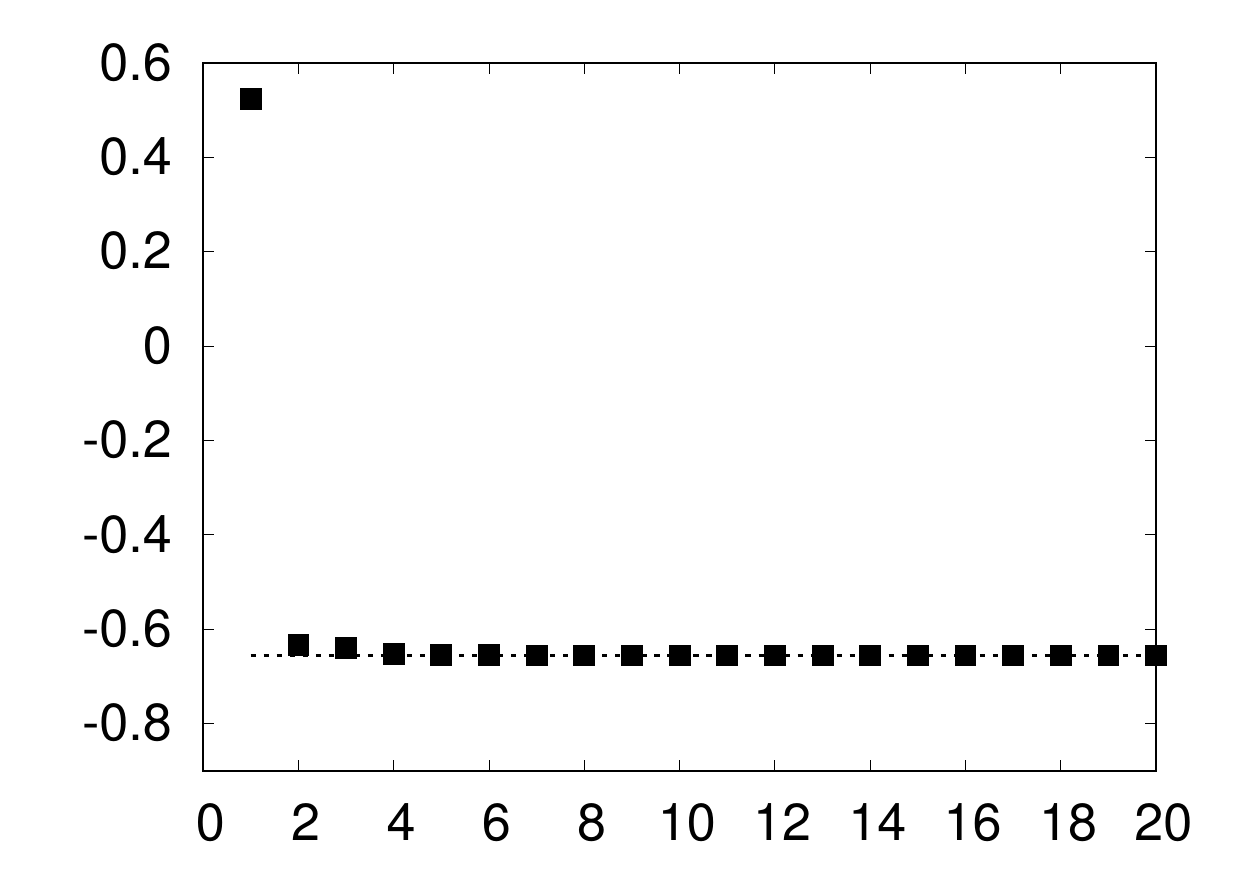}
\caption{$\sigma^z_1\sigma^z_3$}
\end{subfigure}
\begin{subfigure}[b]{0.3\textwidth}
\centering
\includegraphics[scale=0.35]{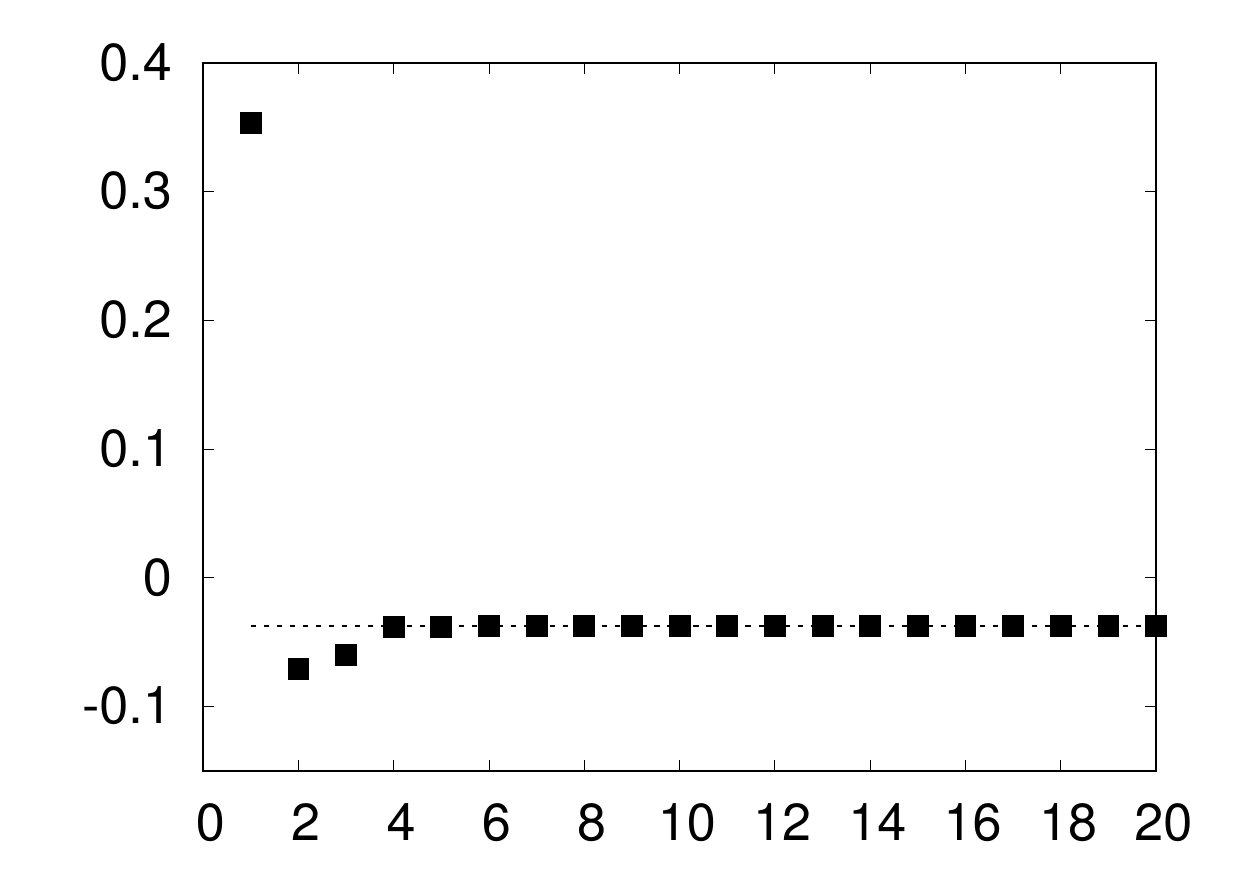}
\caption{$\sigma^z_1\sigma^z_4$}
\end{subfigure}
  \caption{Evaluation of a few local observables within the truncated
    GGE for  the quench from the 4-site DW state. The correlators $\sigma^z_1\sigma^z_a$,
    $a=2,3,4$ are plotted as a function of the truncation index
    $n$. At each truncation step a total of $N_sN_d=n^2$ charges are
    included in the tGGE.
The horizontal lines show the prediction of the full microcanonical GGE.
    The value of the anisotropy is $\Delta=3$.}
  \label{fig:corrDW}
\end{figure}

\begin{figure}
  \centering
\begin{subfigure}[b]{0.3\textwidth}
\centering
\includegraphics[scale=0.35]{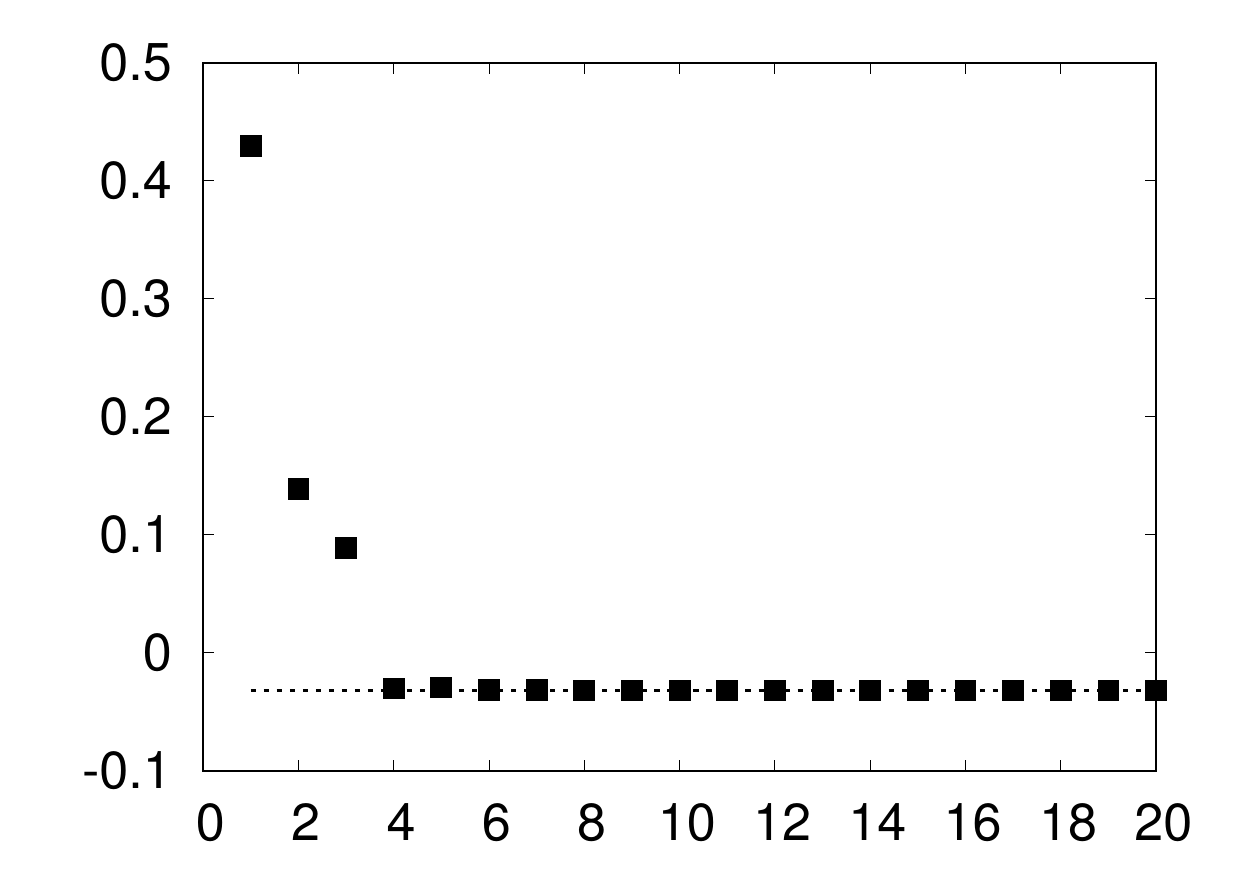}
\caption{$\sigma^z_1\sigma^z_2$}
\end{subfigure}
\begin{subfigure}[b]{0.3\textwidth}
\centering
\includegraphics[scale=0.35]{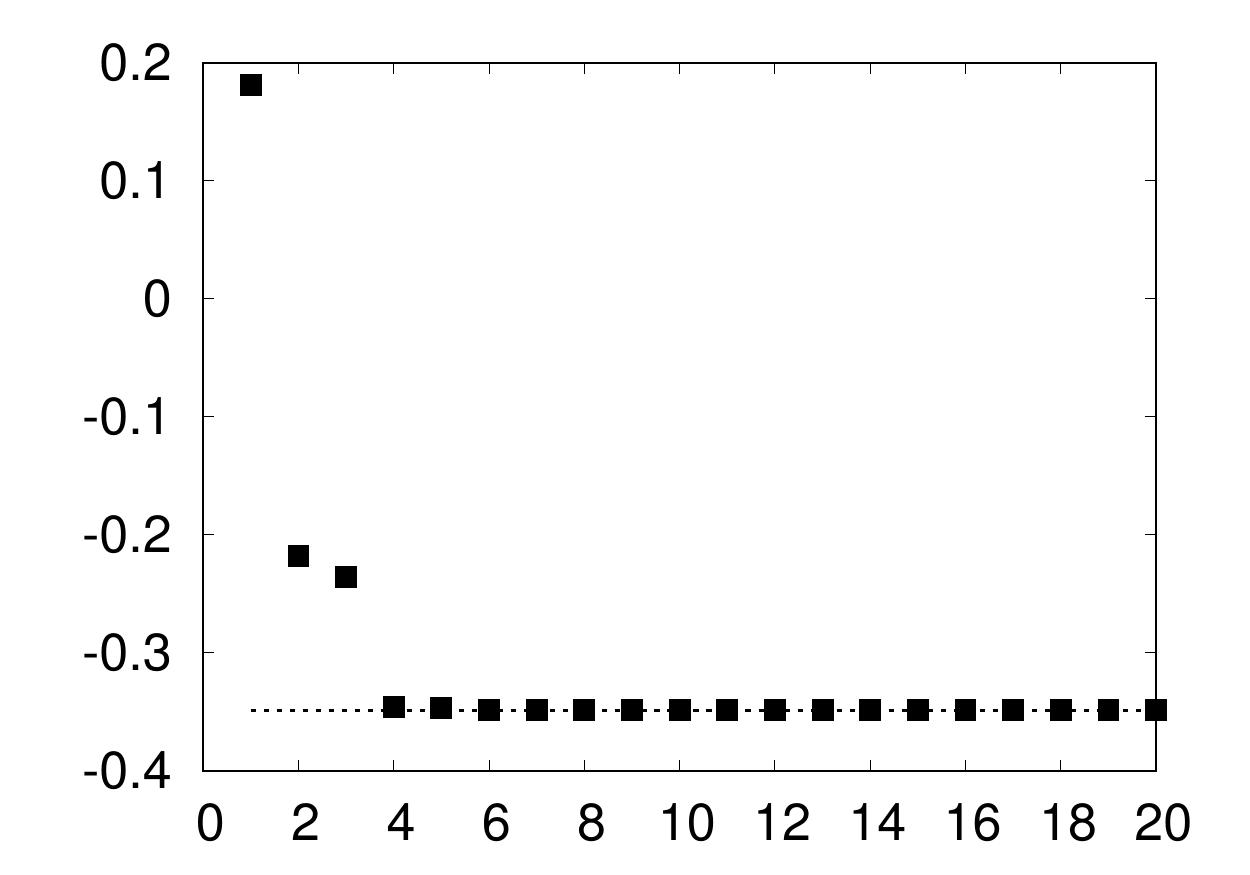}
\caption{$\sigma^z_1\sigma^z_3$}
\end{subfigure}
\begin{subfigure}[b]{0.3\textwidth}
\centering
\includegraphics[scale=0.35]{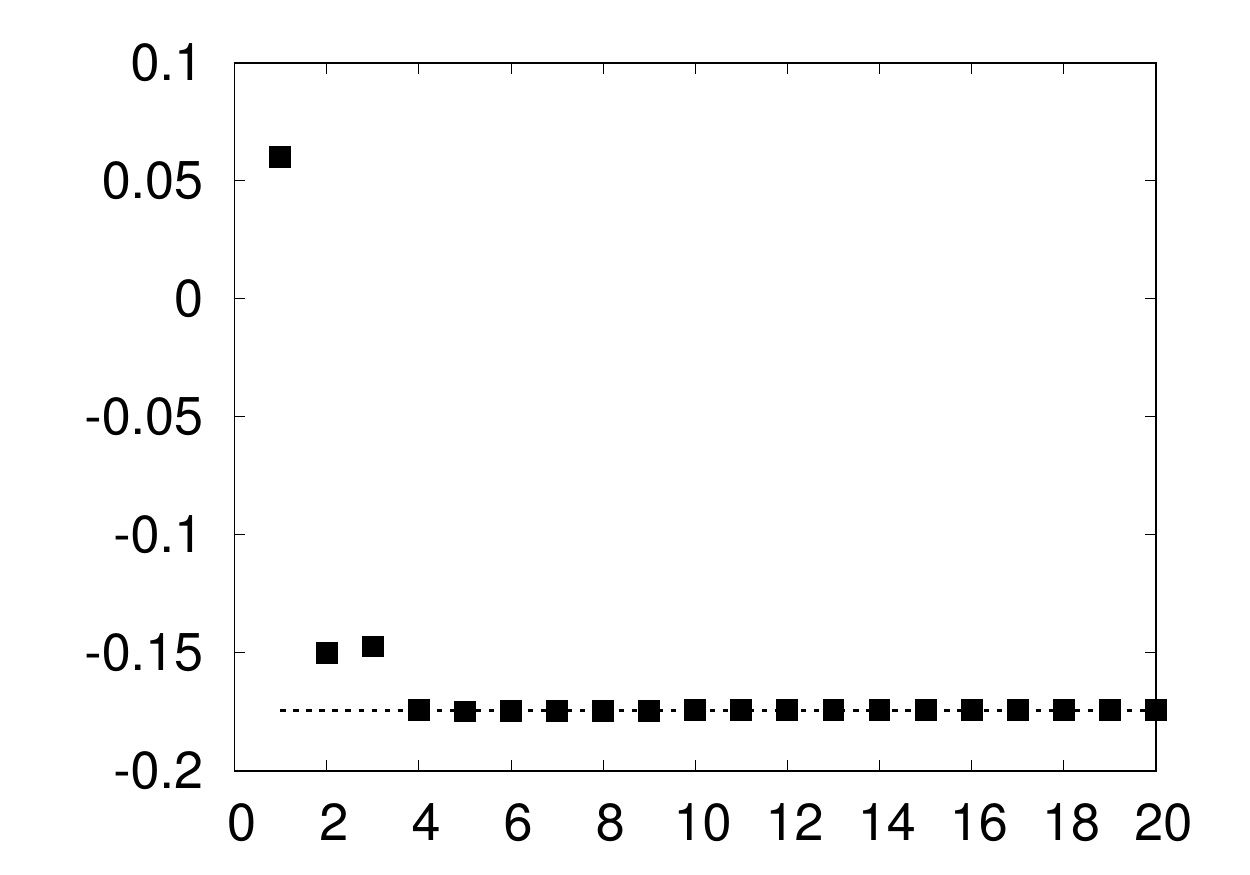}
\caption{$\sigma^z_1\sigma^z_4$}
\end{subfigure}
  \caption{Evaluation of a few local observables within the truncated
    GGE for  the quench from the 4-site generalized dimer state. The correlators $\sigma^z_1\sigma^z_a$,
    $a=2,3,4$ are plotted as a function of the truncation index
    $n$. At each truncation step a total of $N_sN_d=n^2$ charges are
    included in the tGGE.
The horizontal lines show the prediction of the full microcanonical GGE.
    The value of the anisotropy is $\Delta=3$.}
  \label{fig:corr3}
\end{figure}

Figures \ref{fig:corrDW} and \ref{fig:corr3} show the correlators
evaluated within the tGGE's for the DW and generalized dimer states,
respectively. In both cases a fast convergence is observed. A
remarkably good agreement with the full microcanonical ensemble is
achieved already for $n=4$, which corresponds to adding $16$ charges
to the tGGE.

\section{Conclusions}

In this work we re-considered the question whether a canonical
Generalized Gibbs Ensemble can describe steady states after quenches
in the XXZ chain.
Instead of requiring the definition of a GGE including
all charges right from the start, we considered truncated GGE's with
a finite number of discrete charges. Our goal was to show that the
post-quench states can be approximated with arbitrary precision.
To this order we constructed sequences of
truncated GGE's, such that at each step only a finite number
of ultra-local and quasi-local operators are included and the
post-quench Bethe root densities (and therefore all local physical
quantities) are exactly reproduced in the
infinite truncation limit.

Our construction has a number of surprising properties.
The GGE's are built using the canonical set of discrete charges, and
we have shown that the associated Lagrange-multipliers are not well
defined. In practice this means that as we add more and more
charges to the GGE, the values of the first few Lagrange-multipliers
need to be changed as well, and typically they diverge in the infinite truncation
limit (except the first coefficients for each spin/string
index).
This means that they can not be considered physical state functions.
The second striking property is that it is possible to omit a finite number of charges from
the GGE, and the steady states can be reproduced nevertheless,
although with a slower convergence of the tGGE.

Both of these problems can be overcome by constructing a new set of
conserved operators, which is linearly related to the original
set. We have shown that if the new set is chosen properly, the
associated Lagrange-multipliers become well defined state
functions, and they do not change as we add more and more
operators. Conversely, none of the new charges can be omitted from the
GGE. In this sense, the new set of charges plays a similar role as the
Fourier modes of a free system.

These unexpected behaviors can be traced back to known mathematical
facts about infinite dimensional vector spaces.  Our initial goal was to approximate
the exponent of the GGE density matrix using a discrete set of
operators. In practice we have shown that the linear span  of the canonical conserved charges is dense in
the desired subspace of operators. 
However, in an infinite dimensional Banach space a
set of vectors can be dense even when it is not a Schauder
basis, i.e. when the expansion coefficients are not necessarily
well defined. As an effect, the individual coefficients can oscillate or diverge even when the overall
approximation is improving. And it is possible that some vectors can be omitted
from the set such that the linear span will remain dense. However,
after performing a Gram-Schmidt orthogonalization using 
an adequate inner product, the new set of vectors is guaranteed to be a
Schauder basis: all expansion coefficients will be unique and none of
the basis elements can be omitted from the set. We remark that in our construction
the inner product for the Gram-Schmidt orthogonalization is not
related to the original Hilbert space structure of the operators,
instead it is derived from the Bethe Ansatz solution of the model (the
norm is defined as the usual $L_2$ norm of the source functions for the TBA equations).

The physical conclusions to be drawn from our work are as follows. First of all,
we would like to stress the physical relevance of the truncation of the GGE.
In any physical situation one deals with a finite system, where only a
finite number of charges are available.
Moreover, in Section \ref{examples} we have shown that for given
 initial states it is enough to include only $\simeq 100$ charges,
and all local correlators can be reproduced with remarkably good
precision. Therefore, the truncated GGE has strong predictive
power. On the other hand, we have also shown that if the set of conserved
charges is chosen properly, all Lagrange-multipliers are finite,
well defined, and they can be calculated using simple numerical
procedures. Therefore, they are true physical state functions,
  that characterize the system.

It is important to compare our approach and results to those of \cite{enej-gge}.
The construction of \cite{enej-gge} uses the particle number
operators, which  involve through formulas \eqref{rhorho} and 
\eqref{rhofromX} the generating functions of the higher charges evaluated at
an infinitesimally small distance from the boundary of the physical
strip.
The resulting operators are quasi-local for any finite value of the
regulator, but they lose quasi-locality as this regulator is taken to
zero.
In contrast, our construction uses a discrete set of quasi-local
operators such that the range of the operators is increased
gradually. Nevertheless, the exponent of the tGGE
is expected to lose the quasi-locality property 
 in the infinite truncation limit. We stress that the
 physical observables are exactly reproduced by both approaches, and
 the difference lies only in the choice of operators to approximate the exact
 GGE. Our method is built on the traditional requirement of statistical
 physics, namely that Gibbs or Generalized Gibbs Ensembles should be
 built using sufficiently local quantities, whereas the
 method of \cite{enej-gge} provides a close connection to the particle
 picture, and to the GGE in free theories.
We believe it would be useful to understand better the relation
between the two approaches, and in particular the connection between
the infinite truncation limit on one
 side, and the zero-regulator limit on the other.  

\bigskip
  
Our results open the way for a number of further interesting questions.
  
It would be interesting to investigate which charges are the most
relevant for the truncated GGE. In the present work we applied a
simple procedure where at each truncation step we included the
first $n$ charges from the first $n$ spin/string families. However,
other schemes could be considered as well. This might shed light on
how many discrete charges are actually needed to achieve a certain accuracy of
the tGGE.

It would be useful to investigate quenches where the initial
state is not a simple product state, for example the ground state of
an other XXZ Hamiltonian.
Our general derivation shows that
the construction works for these states as well, and it would be
interesting to see how the number of the required charges depends on
the initial state.

Finally, it would be interesting to study truncated GGE's in other integrable
models too.
The essence of our construction, namely the technique of
approximating the source terms of a TBA using a discrete set of
charges is not limited  to the XXZ chain. We expect similar
behavior for other integrable models too, irrespective of their particle
content. Note, that in our calculations the string structure of the TBA did not
play a crucial role: the approximation procedure worked for each
string index separately, resulting in the same linear transformation
\eqref{tildeQdef} for the charge families. Therefore, our construction
is expected to work for various models which admit a TBA-like solution.

We hope to return to these questions in further research.

\vspace{1cm}
{\bf Acknowledgments} 

We are grateful to Jean-S\'ebastien Caux, Enej Ilievski, M\'arton Kormos, Izabella Lovas,
Marko Medeniak, Lorenzo Piroli, Frank Pollmann, Toma\v{z} Prosen, 
Eoin Quinn, Junji Suzuki and G\'abor Tak\'acs  for useful discussions. 
We are also thankful to Lorenzo Piroli for a careful reading of the manuscript.
 The proof of Theorem \ref{remling} was given by Christian Remling on 
Math.Stackexchange \cite{876492}, and we are grateful to him for
letting us use it.
B. P. acknowledges support from the ``Premium'' Postdoctoral
Program of the Hungarian Academy of Sciences, and the NKFIH grant no. K119204.
E. V. acknowledges support from the ERC under Starting Grant 279391 EDEQS.
M. W. acknowledges support from the NKFIH grants no. K105149 and SNN118028.

\bigskip

\addcontentsline{toc}{section}{References}
\bibliography{tGGEcikk}
\bibliographystyle{utphys}

\end{document}